\def\aprge{\buildrel > \over {_{\sim}}}

\def\Dm {\Delta m^2}
\def\stheta {\sin^2 2\theta}
\newcommand{\cm}{\rm \,cm}

\newcommand{\m}{\rm \,m}
\newcommand{\nm}{\rm \,nm}

\newcommand{\s}{\rm \,s}

\newcommand{\sr}{\rm \,sr}

\newcommand{\mg}{\rm \,mg}

\newcommand{\g}{\rm \,g}
\newcommand{\hg}{\rm \,hg}
\newcommand{\MeV}{\rm \,MeV}

\newcommand{\GeV}{\rm \,GeV}
\newcommand{\TeV}{\rm \,TeV}

\newcommand{\MHz}{\rm \,MHz}

\def\ne {\nu_e}
\def\nm {\nu_\mu}
\def\nt {\nu_\tau}

\newcommand{\anue}{$\bar{\nu}_e$}

\def\nmnt {\nu_\mu \longleftrightarrow \nu_\tau}
\def\nmns {\nu_\mu \longleftrightarrow \nu_{sterile}}
\newcommand{\lsim}{\lower .5ex\hbox{$\buildrel < \over {\sim}$}}
\newcommand{\gsim}{\lower .5ex\hbox{$\buildrel > \over {\sim}$}}

\newcommand{\stitolo}{The MACRO Experiment at Gran Sasso}
\newcommand{\sautori}{G.~Giacomelli and A.~Margiotta}
\newcommand{\sistituzioni}{Dipartimento di Fisica dell'Universit\`{a} di Bologna and INFN, 40127 Bologna, Italy}
\newcommand{\sabst}{In this overview of the MACRO experiment we recall the structure of the detector and discuss several physics topics: atmospheric neutrinos and neutrino oscillations, high energy neutrino astronomy, searches for WIMPs and for low energy stellar gravitational collapse neutrinos, stringent upper limits on GUT magnetic monopoles, high energy downgoing muons, primary cosmic ray composition and shadowing of primary cosmic rays by the Moon and the Sun.}
\documentclass[12pt,a4paper]{article}
\usepackage{epsfig}                     
\newcommand{\titolo}{\begin{center}{\Large\bf{\stitolo}}\end{center}}
\newcommand{\autori}{\begin{center}\sautori\end{center}}
\newcommand{\istituzioni}{\begin{center}\small{\it{\sistituzioni}}\end{center}}
\newcommand{\abst}{\begin{abstract}\sabst\end{abstract}}

\begin{document}
\pagestyle{empty}
\enlargethispage{2cm}
\titolo
\autori
\istituzioni
\begin{center}
\textbf{Charles  Peck - Fest, Caltech,  2005}
\footnote{The meeting was organized by Douglas Michael, who later passed away. Doug was one of the leaders of MACRO.}
\end{center}
\abst
%
\pagestyle{plain}                
%
\section{Introduction}
MACRO was a large area multipurpose underground detector designed to search for rare events and rare phenomena in the penetrating cosmic radiation. It was optimized to look for the supermassive magnetic monopoles predicted by Grand
Unified Theories (GUT) of the electroweak and strong interactions; it could also perform measurements in areas of astrophysics, nuclear, particle and cosmic ray physics. These include in particular the study of atmospheric
neutrinos and neutrino oscillations, high energy neutrino astronomy, indirect searches for WIMPs, studies of various aspects of the high energy underground muon flux (which is an indirect tool to study the primary cosmic ray composition), searches for rare particles that may exist in the cosmic radiation. \\
\indent The experiment was located in Hall B of the underground Gran Sasso Lab and started data taking with part of the apparatus in 1989; it was completed in early 1995 and  was running in its final configuration until the end of 2000. 
The detector had global dimensions of $76.6\times 12 \times 9.3$ m$^3$ and provided a total acceptance of $\sim$10000 $ m^2 sr$ to an isotropic flux of particles. 
The  detector had a modular structure: it was divided into six sections referred to as supermodules.
The average rock overburden was \( \simeq 3700 \) m.w.e., while the minimum was \( 3150 \) m.w.e. This defines the minimum muon energy at the surface as $\sim$1.3 $\TeV$  in order to reach MACRO. 
\textheight24cm       

It may be worth pointing out that all the physics and astrophysics items proposed in the 1984 Proposal were covered and good results were obtained on each of them, even beyond the most rosy anticipations \cite{mac1}-\cite{mac52}.

The collaboration consisted of $\sim$140 physicists and engineers from 6 Italian and 6 US Institutions, 1 Moroccan group and visitors from various developing countries. The complete list of names can be found in Ref. \cite{mac10}, \cite{mac35}, \cite{mac52}. There was a close cooperation with the EASTOP Collaboration which operated an Extensive Air Shower Array at Campo Imperatore.
\newpage      
\section{The Detector}
\begin{figure}
  \begin{center}
  \mbox{ \epsfysize=8cm    \epsffile{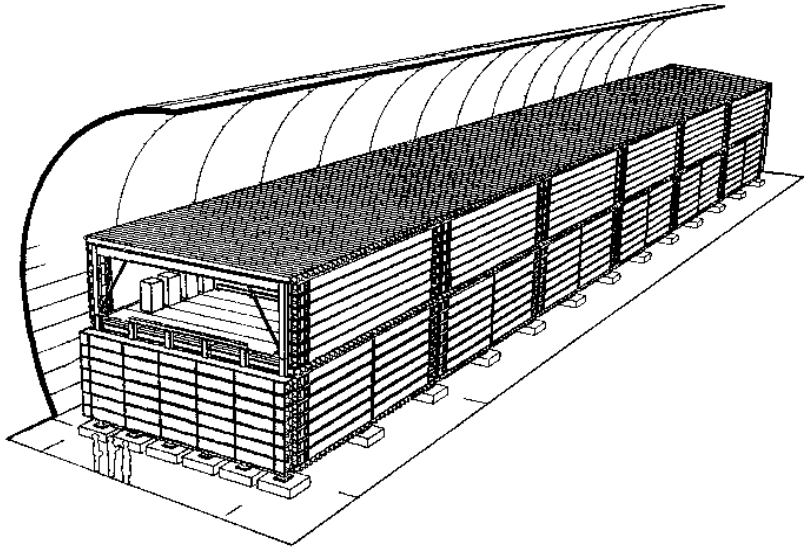} }   
\caption{\label{fig1}\footnotesize {Layout of the MACRO detector which was installed in Hall B of the LNGS.
Overall dimensions of the active part were $76.5\times12\times9.3~{\m}^3$  \cite{mac35}.}
}
\end{center}
\end{figure}

\begin{figure}
  \begin{center}
  \mbox{\epsfysize=8cm \epsffile{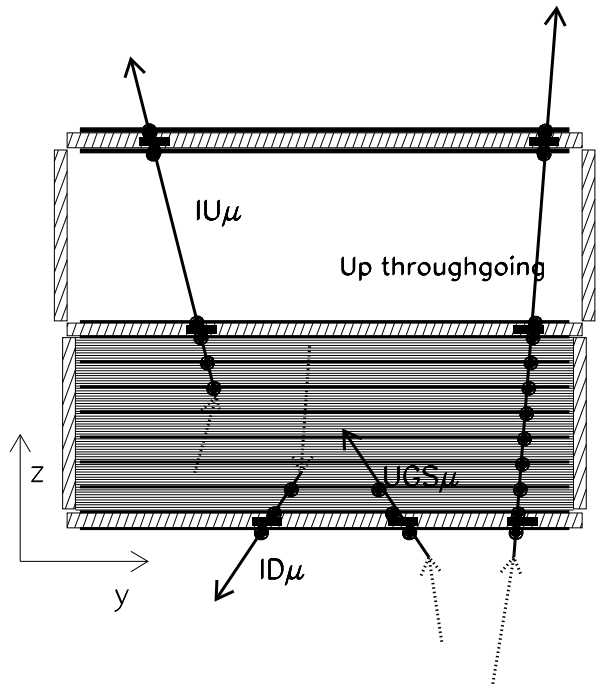} }   
\caption{\label{fig2}\footnotesize {Vertical cross section of the detector and sketch of different event topologies induced by $\nu_\mu $ interactions in or around MACRO. The black points and the black rectangles represent streamer tubes and scintillator hits, respectively. Tracking was performed by the streamer tubes;  the time-of-flight of the muons was measured by the scintillators for {\it Up Semicontained } (Internal upgoing - IU $\mu$) and {\it Upthroughgoing }  events (and also for downgoing muons).}
}
\end{center}
\end{figure}

Redundancy and complementarity were the primary goals in designing the experiment. Since  few events were 
expected, multiple signatures and ability to perform cross checks with various parts of the apparatus were important. The detector was composed of three sub-detectors: liquid scintillation counters, limited streamer tubes and nuclear track detectors \cite{mac16}\cite{mac35}. Each one of them could be used in \lq\lq stand-alone\rq\rq~and in \lq\lq combined\rq\rq~mode. The layout of the experiment is shown in Fig. \ref{fig1}. Notice the division in the \textit{lower}  and in the  \textit{upper} part (this was often referred to as the \textit{Attico}); the inner part of the \textit{Attico} was empty and lodged the electronics. Fig. \ref{fig2} shows a cross section of the apparatus.

{\bf The scintillation subdetector.} Each supermodule contained 77 scintillation  counters, divided into three horizontal planes (bottom, center, and top) and two vertical planes (east and west). The lower part of the north and south faces of the detector were covered by vertical walls with 7 scintillators each; the upper parts of these faces were open to allow access to the readout electronics. 
The active volume of each horizontal scintillator was \( 11.2\times 0.73\times 0.19{\m }^{3} \),
while for the vertical ones it was  \( 11.1\times 0.22\times 0.46{\m }^{3} \).
All scintillator boxes were filled with a mixture of high purity mineral oil (\( 96.4\, \% \)) and
pseudocumene (\( 3.6\, \% \)), with an additional \( 1.44{\g } \)/l of PPO and \( 1.44{\mg } \)/l of bis-MSB wavelength shifters. The horizontal counters were seen by two \( 8^{\prime \prime } \) photomultipliers (PMTs) and the vertical counters by one \( 8^{\prime \prime } \) PMT at each end.  The total number of scintillators was \( 476 \) with a total active mass of \(\simeq  600\, {\textrm{tons}} \). Minimum ionizing muons, when crossing vertically the \( 19 {\cm } \) of scintillator,  released an average energy of \( \simeq 34{\MeV }
\) and were measured with a timing and longitudinal position resolution of \( \simeq 500\, {\textrm{ps}} \) and \( \simeq 10{\cm } \), respectively.

The scintillation counters were equipped with specific triggers for rare particles, muons and low energy neutrinos from stellar gravitational collapses. The Slow Monopole Trigger (SMT) was sensitive to MMs with velocities from  \( 10^{-4}c \) to \( 10^{-2}c \), the Fast Monopole Trigger (FMT)  to poles with velocities from  \( 5\times10 ^{-3}c \) to \( 5\times10 ^{-2}c \), the Lightly Ionizing Particle trigger was sensitive to fractionally charged particles, the Energy Reconstruction Processor (ERP) and {}``CSPAM{}'' were primarily muon triggers (but used also for relativistic monopoles)
and the gravitational collapse neutrino triggers (the Pulse Height Recorder and Synchronous Encoder --PHRASE-- and the ERP) were optimized to trigger on bursts of low energy events in the liquid scintillators. The scintillator system was complemented by a \( 200{\MHz } \) waveform digitizing (WFD) system used in rare particle searches, and in any occasion 
where knowledge of the PMT waveform was useful.

{\bf The streamer tube subsystem.} \textit{The lower part} of the detector contained 10 horizontal planes of limited streamer tubes, the middle 8 of which were interleaved by 7 rock absorbers (total thickness \( \simeq 360{\g }{\cm }^{-2} \)). These set a \( \simeq 1{\GeV } \) energy threshold for muons vertically crossing the lower part of the
detector. At the top of the \textit{Attico} there were 4 horizontal streamer tube planes, 2 above and 2 below the top scintillator layer. On each lateral wall 6 streamer tube planes sandwiched the corresponding vertical scintillator plane (3 streamer planes on each side). Each tube had a \( 3\times 3{\cm }^{2} \) cross section and was \( 12{\m } \) long. The total number of tubes was \( 50304 \), all filled with a gas mixture of \( He \) (\( 73\, \% \)) and n-pentane
(\( 27\, \% \)). They had \( 100{\mu } \) \textit{Cu/Be} wires and stereo pickup strips at an angle of \( 26.5^{\circ } \). The tracking resolution of the streamer tube system was \( \simeq 1{\cm } \), corresponding to an angular accuracy of \( \simeq 0.2^{\circ } \) over the 9.3 m height of MACRO. The real angular resolution was limited to \( \simeq 1^{\circ } \) by  multiple Coulomb scattering of muons in the rock above the detector. The streamer tubes were read by \( 8 \)-channel cards (1 channel for each wire) which discriminated the signals and sent the analog information (time
development and total charge) to an ADC/TDC system (the QTP). The signals were used to form 2 different chains (Fast and Slow) of  pulses, which were the inputs for the streamer tube Fast and Slow Particle Triggers. In the 11 years of operation only 50 wires were lost.

{\bf The nuclear track subdetector} was deployed in three planes, horizontally in the center of the lower section and vertically on the East and North faces. The detector was divided in \( 18126 \) modules, which could be individually extracted and substituted.  Each module (\( \sim 24.5\times 24.5\times 0.65 {\cm }^{3} \) ) was composed of three layers of CR\( 39 \), three layers of Lexan and \( 1\, {\textrm{mm}} \) Aluminium absorber to stop nuclear
fragments.

{\bf The Transition Radiation Detector (TRD)} was installed inside  the \textit{Attico}. It was composed of 3  modules (overall dimensions \( 6\times 6\times 2{\m }^{3} \)) and it was made of \( 10{\cm } \) thick polyethylene foam radiators and proportional counters filled with \( Ar \) (\( 90\, \% \)) and \( CO_{2} \) (\( 10\, \% \)). The TRD  measured  muon energies in the range \( 100{\GeV }<E_\mu<930{\GeV } \); muons of higher energies were counted. 
 
\begin{figure}
  \begin{center}
  \mbox{\epsfysize=6cm    \epsffile{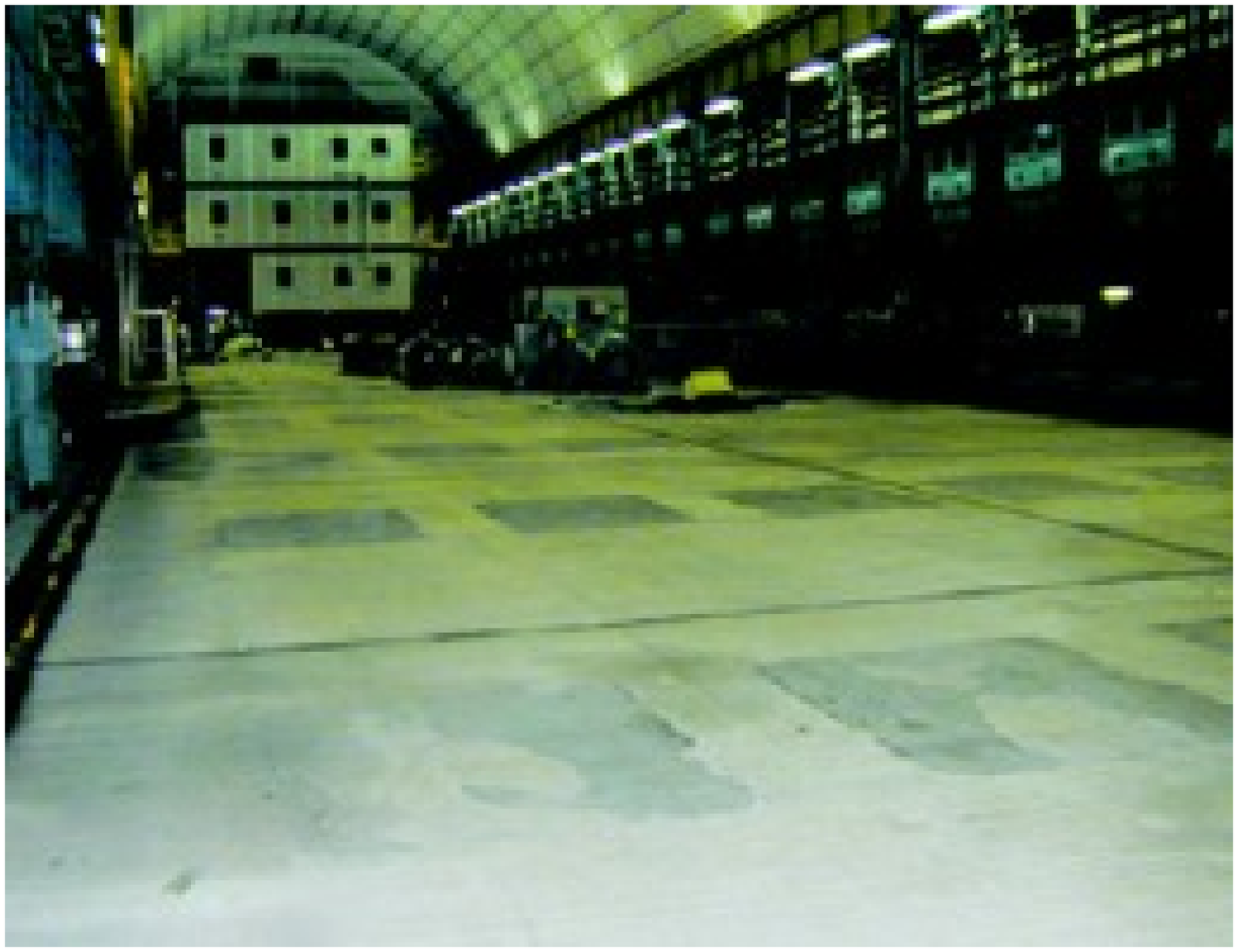}  
  \epsfysize=6cm    \epsffile{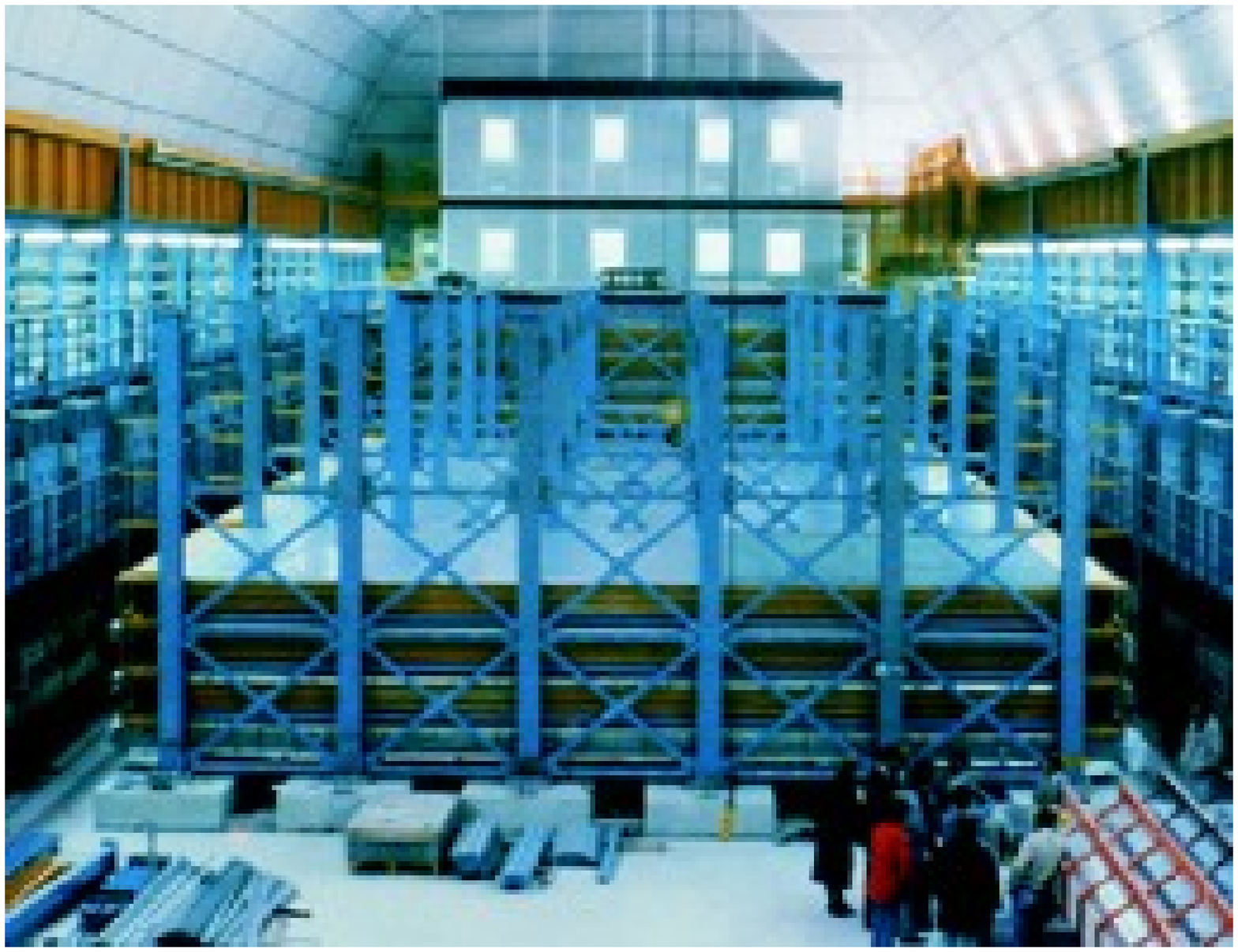}}
{\small \hskip 6.0 truecm (a) \hskip 8.5 truecm (b)}
  \mbox{
\epsfysize=6cm    \epsffile{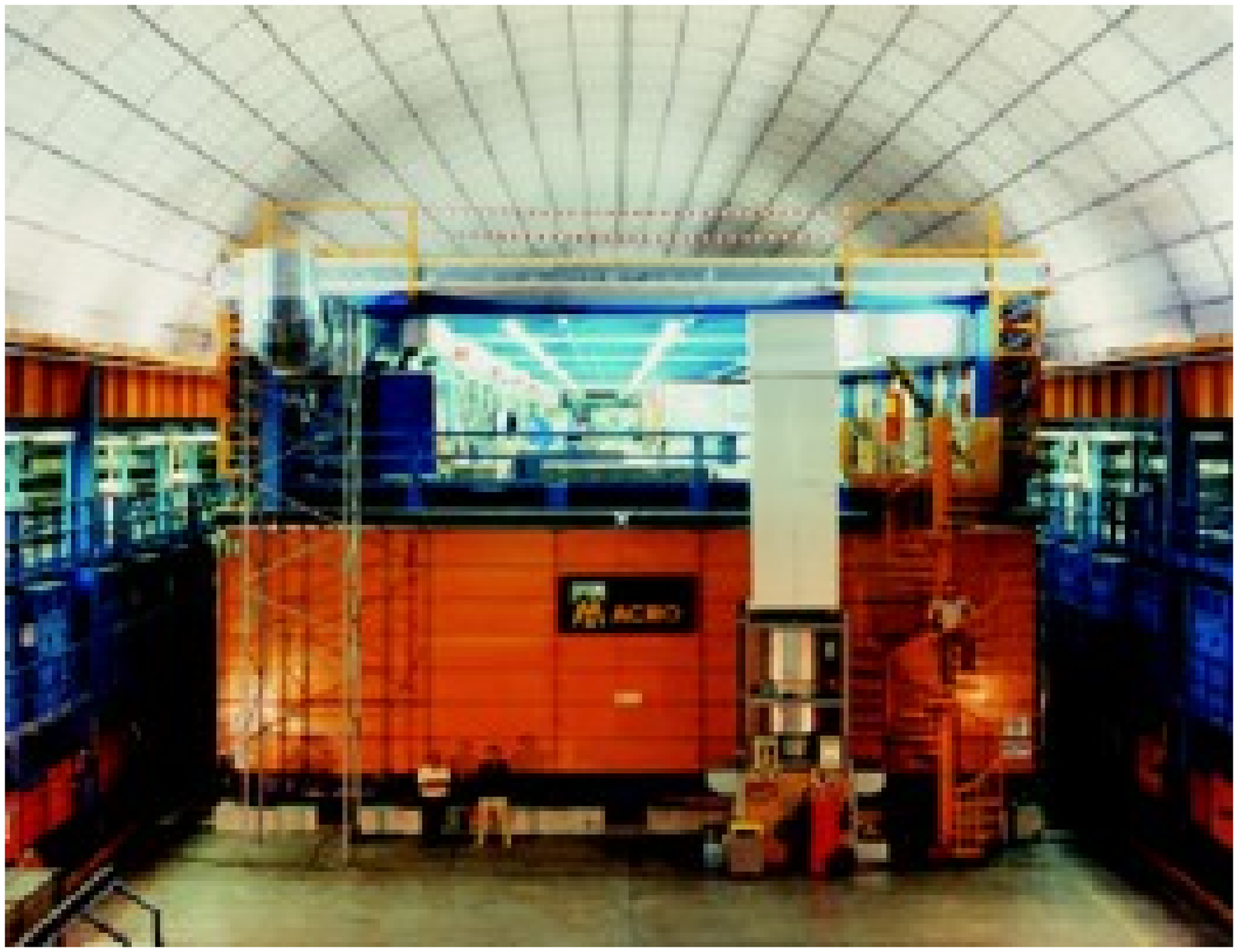}  
  \epsfysize=6cm    \epsffile{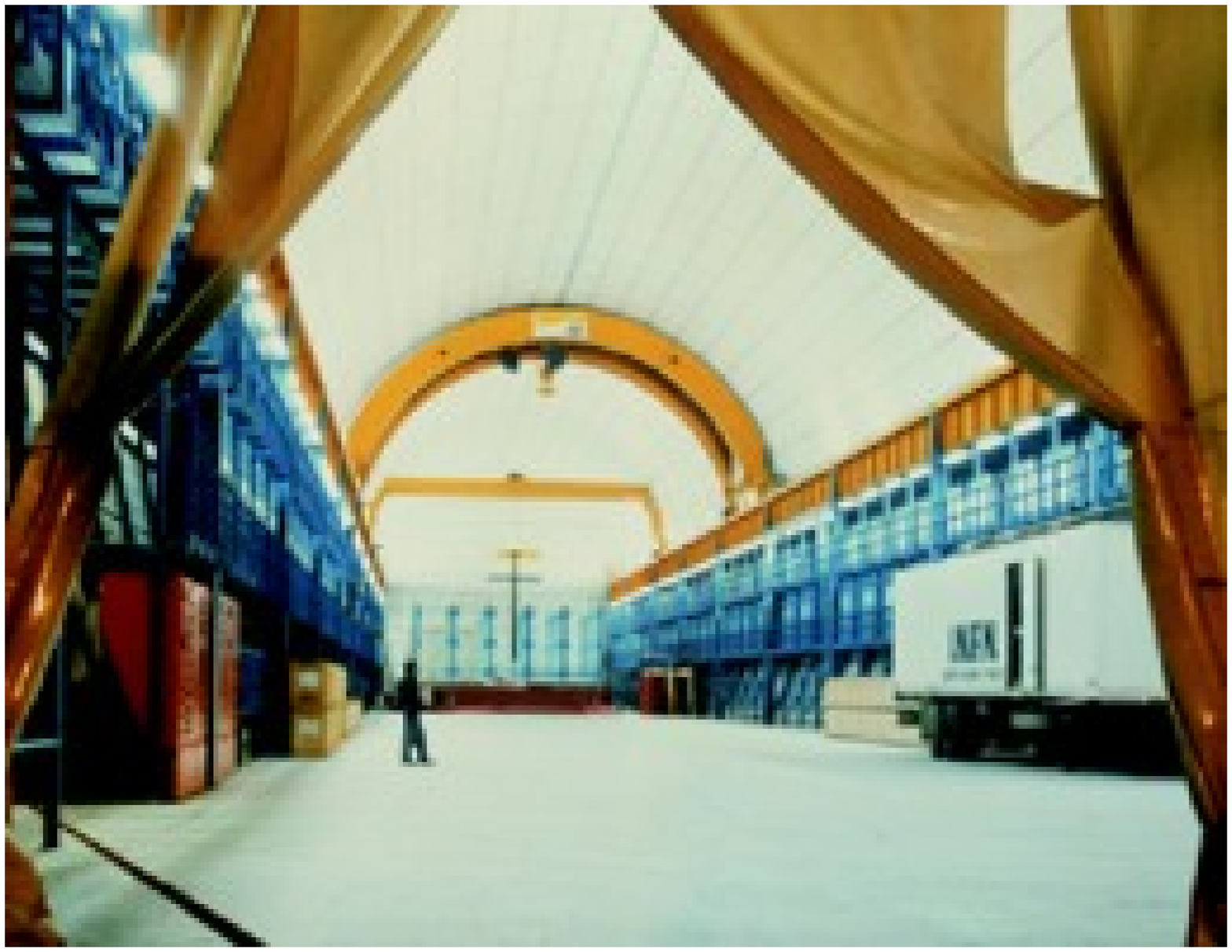}}
{\small \hskip 6.0 truecm (c) \hskip 8.5 truecm (d)}
 \end{center}
  \caption {\label{fig3}\footnotesize  {Hall B: (a) in 1987 just before starting construction; (b) in 1990 when the first lower supermodule was taking data while the second and  third were under construction; (c) in 1995 when the completed MACRO detector started data taking (safety stairs and a ventilation system were added later); (d) Hall B empty in 2001. }}
\end{figure}

\begin{figure}
 \begin{center}
 \hspace{1.cm}
  \mbox{ \epsfysize=3.6cm
         \epsffile{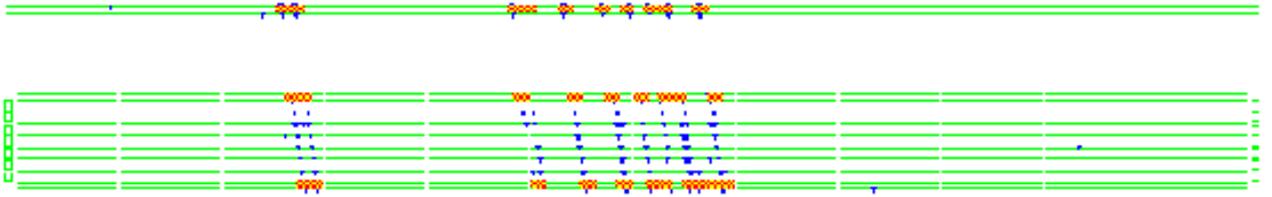} }
 \end{center}
\caption {\label{fig4}\footnotesize { MACRO Event Display: lateral view of a group of 11 downgoing  muons.}}
\end{figure}

Fig. \ref{fig3} shows four photographs of  Hall B taken from its south side. Fig. \ref{fig4} shows a {}``group{}'' of 11 downgoing muons.
\section{Atmospheric neutrino oscillations}
A high energy primary cosmic ray (CR), proton or nucleus, interacts in the upper atmosphere producing  a large number of charged pions and kaons, which decay yielding  muons and muon neutrinos; also the muons decay yielding \( \nu _{\mu } \) and \( \nu _{e } \). The ratio of the numbers of \( \nu _{\mu } \) to \( \nu _{e } \)  is $\simeq$  2 
and $N_{\nu}/N_{\overline\nu} \simeq 1$. Atmospheric  neutrinos are produced in a spherical surface at  10-20 km above ground and they proceed towards the earth. 

If neutrinos have non-zero masses, one has to consider the {\it weak flavour eigenstates} $~\ne,~\nm,~\nt$ and the {\it mass eigenstates} $~\nu_1,~\nu_2,~\nu_3$.  The  flavour eigenstates are linear combinations of the mass eigenstates. For 2 flavour  ($\nm,~\nt$) and 2 mass eigenstates $(\nu_2,~\nu_3)$ 

\begin{equation}
\left\{ \begin{array}{ll}
      \nm =~\nu_2 \cos\ \theta_{23} + \nu_3 \sin\ \theta_{23} \\
      \nt=-\nu_2\sin\ \theta_{23} + \nu_3\cos\ \theta_{23} 
\end{array} 
\right. 
\end{equation} 
\noindent where $\theta_{23}$ is the mixing angle. The survival probability of the $\nm$ "beam" is
\begin{equation}
P(\nm \rightarrow \nm) = 1- \sin^2 2\theta_{23}~\sin^2 \left( 
{\frac{1.27 \Dm \cdot L} {E_\nu}} \right)
\end{equation}
\noindent where $\Dm=m^2_3-m^2_2$, $L$ is the distance travelled by the 
$\nu$ from production to detection. 

Atmospheric neutrinos are well suited for the study
of neutrino oscillations, since they have energies $E_\nu$ from a fraction of GeV up 
to more than 100 GeV and they travel distances $L$ from few tens of km 
up to 13000 km; thus $L/E_\nu$ ranges from $\sim 1$ km/GeV to $\sim 10^5$ 
km/GeV.  They are particularly useful to study oscillations
for small $\Dm$, while matter effects can be studied with their high
energy components. 

The early water Cherenkov detectors and the tracking calorimeters
measured $\nm$ 
and $\ne$ charged current (CC) interactions. The results were
expressed in terms of the double ratio $R^\prime =R_{obs}/R_{MC}$, where
$R_{obs} = (N_{\nm} / N_{\ne})_{obs}$ is the ratio of observed 
$\mu$ and $e$ events and $R_{MC} = (N_{\nm}/ N_{\ne})_{MC}$ 
is the same ratio for Monte Carlo (MC) events. The $R^\prime$ ratios from IMB \cite{imb} and Kamiokande \cite{kamioka} were smaller than expectations, while 
NUSEX \cite{nusex}, Frejus \cite{frejus} and Baksan \cite{Baksan} did not find any deviation.
Later, the Soudan 2 tracking and shower calorimeter detector confirmed the 
anomaly in the $\nm/\ne$ double ratio for contained events \cite{soud2}. MACRO reported 
in 1995 a measurement of upthroughgoing muons from $\nm$ of  $\langle E_\nu\rangle$$\sim$50 GeV, in which there was an anomalous zenith distribution and a deficit in the total number of observed upgoing muons \cite{mac17}. 
In 1998 Soudan 2, MACRO and SuperKamiokande (SK) provided strong indications in favour of $\nmnt$ oscillations \cite{soud2_b}\cite{mac25}\cite{sk}\cite{macro}.  Later new results were presented by the 3 experiments and by others \cite{k2k}. \\
\indent MACRO detected upgoing $\nm$'s via CC interactions $\nm \rightarrow \mu$; upgoing muons were identified with the streamer tube system (for tracking) and the scintillator system (for time-of-flight measurement). The events were classified in different categories:\\
{\bf \boldmath Upthroughgoing muons ($E_\mu > 1$ GeV)} come from interactions in the rock below the detector of $\nm$ with $\langle E_\nu$$\rangle$$\sim$50 GeV. The MC uncertainties arising from the neutrino flux, cross section and muon propagation on the expected flux of upthroughgoing muons were estimated to be $\sim$17$\%$; this systematic error  is 
mainly a scale uncertainty.

\begin{figure}
\begin{center}
\mbox{\hspace{-1.5cm}\epsfig{figure=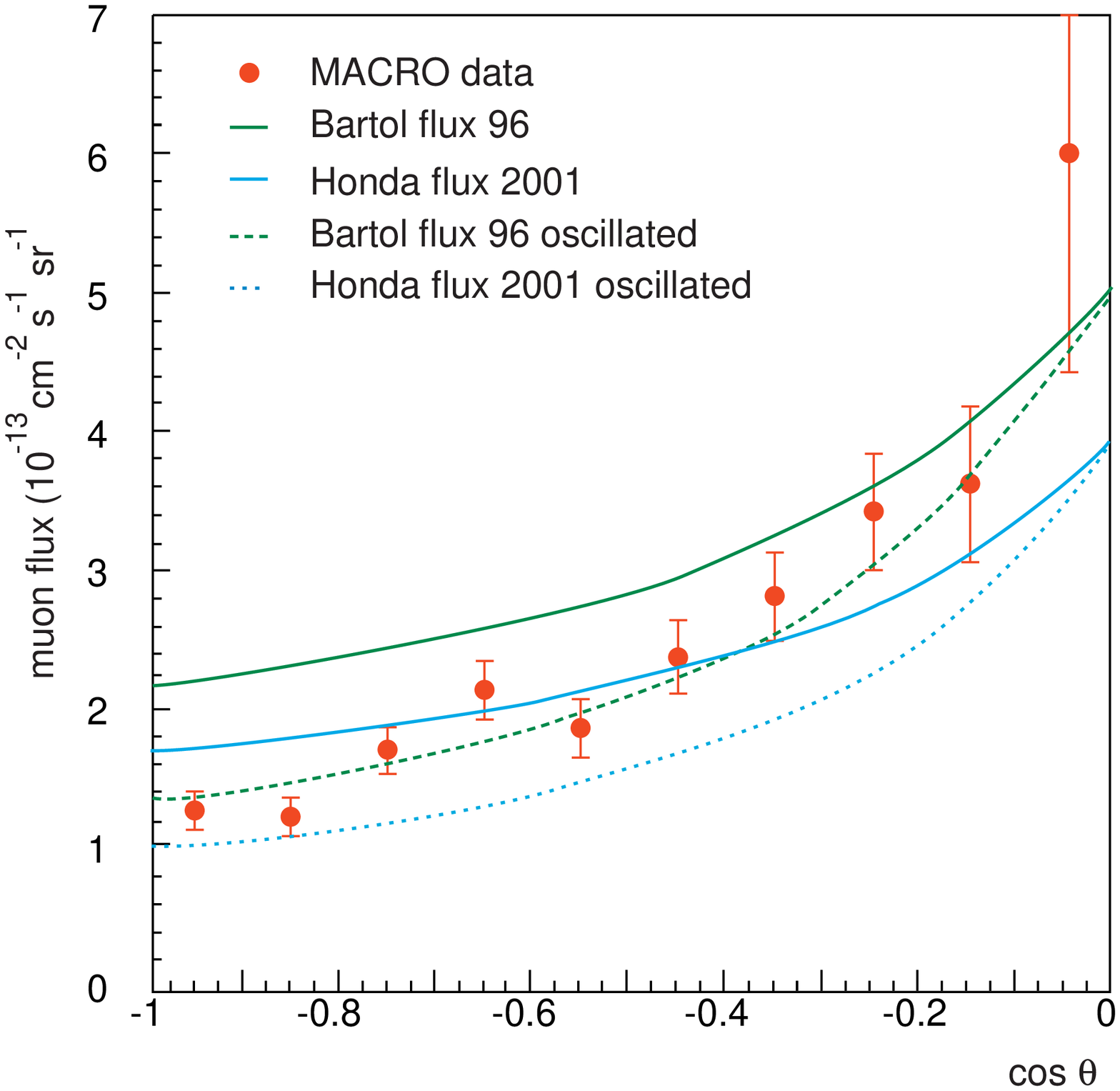,height=5.5cm}
\epsfig{figure=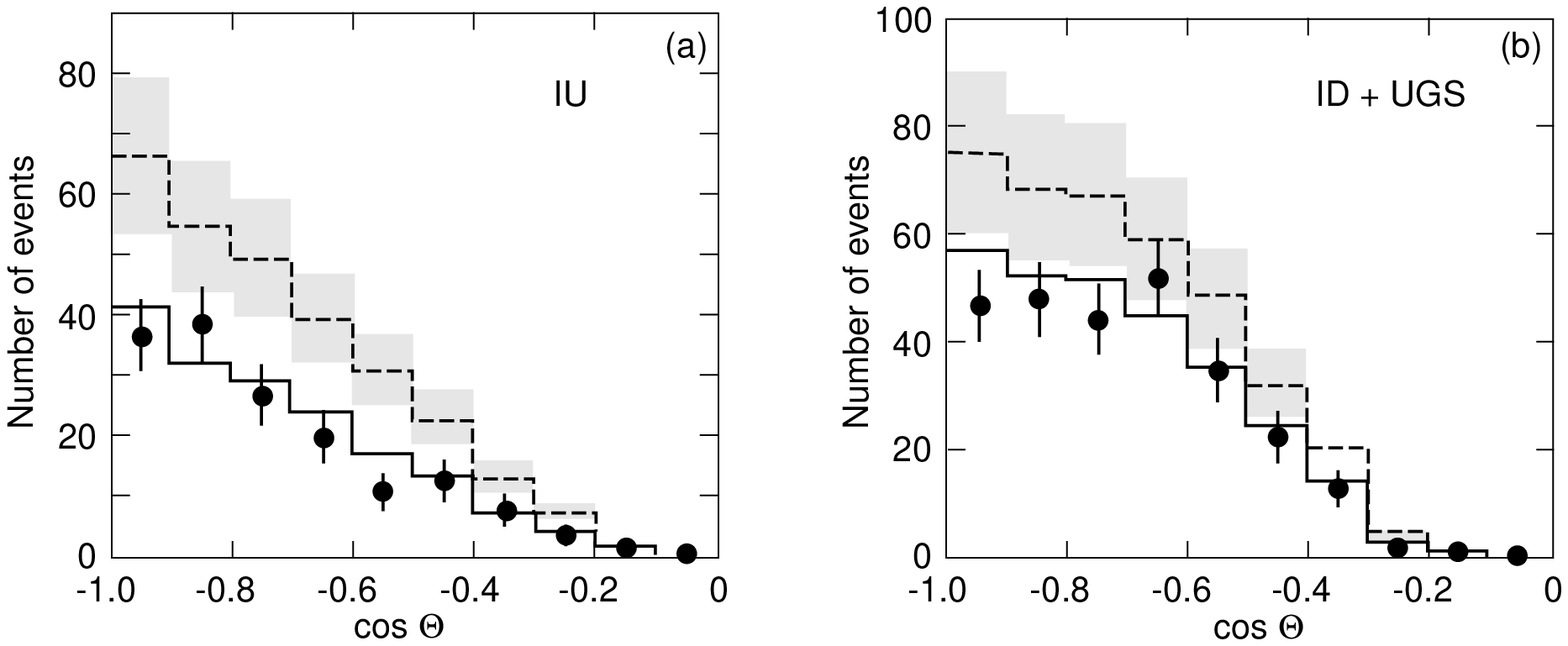,height=5.5cm}
}
{\footnotesize (a) \hspace {5cm} (b) \hspace {5cm} (c)}
\caption { \footnotesize {(a) Comparison between the zenith distribution of the MACRO upthroughgoing muons and the oscillated MC predictions given by Bartol96 (solid curve), HKKM01 (dash-dotted line), FLUKA fitted to the new CR measurements (dashed curve) and FLUKA with the old CR fit (dotted curve). Zenith distributions  (b) for IU and  (c) for ID+UGS  events (black points) compared with the no oscillation Bartol96 MC (dashed line with a scale error band) and with the $\nmnt$ predictions with $\Dm=2.3 \cdot 10^{-3}$ eV$^2$ and maximal mixing.}}
\label{fig:cosze}
\end{center}
\end{figure}

In order to verify that different flux simulations affect the zenith distribution at the level of only a few percent (while there is an effect of the order of $\sim 25\%$ on the event rates) MACRO compared data with the predictions of the Bartol96 \cite{Bartol96}, FLUKA \cite{FLUKA} and HKKM01 \cite{honda01} MCs, see fig. \ref{fig:cosze}a. The shape of the angular distribution and the absolute value strongly favour neutrino oscillations with $\Dm = 0.0023$ eV$^2$ and maximum mixing.
The absolute values of the MACRO upthroughgoing muon data are  $25\%$ higher than those predicted by the  FLUKA and HKKM01 MC, while the shapes of the oscillated and non oscillated angular distributions from the different MCs agree within 5\%.

A similar situation is found in the SK data \cite{sk}. The electron-like events were in agreement with the HKKM95 \cite{honda96} MC predictions in absence of oscillations, while they are higher than the HKKM01 \cite{honda01} non oscillated MC. For  muon-like events, the new MC predictions are low for the SK data, especially for upthroughgoing muons. Previous comparisons between the SK muon data and the HKKM95 predictions had shown a global deficit of events and a zenith distribution shape in agreement with $\nmnt$ oscillations \cite{sk}.

The difference between the new and old MC predictions is probably due to the use of a new fit of the cosmic ray data \cite{gaisser}. Recent results by the L3C and BESS experiments \cite{L3Bess} on the primary cosmic ray fit show good agreement with the Bartol96 and HKKM95 predictions and a disagreement with the new fit of the cosmic ray data \cite{gaisser}. \\
{\bf Low energy events.} {\it Semicontained upgoing muons} (IU) come from
$\nm$ interactions inside the lower apparatus. {\it Up stopping muons} (UGS)
are due to external $\nm$ interactions yielding upgoing muons stopping in the 
detector; the {\it semicontained downgoing muons} (ID) are due to downgoing 
$\nm$'s with interaction vertices in the lower detector; the lack
of time information prevents to distinguish between the two subsamples.
An almost equal number of UGS and ID events is expected. The average 
parent neutrino energy for all these events is 2-3 GeV. 
The  angular distributions are compared with  MC
predictions without oscillations in   Figs. \ref{fig:cosze}b,c. Our low
energy data show a uniform deficit 
over the whole angular distribution with respect to the Bartol96 predictions, thus favouring $\nu$ oscillations.
\begin{table}[ht]
\begin{center}
\begin{tabular}{@{}cccc@{}}
\hline
{} &{} &{} &{}\\[-1.5ex]
{} & Events & MC$_{\mbox{no~osc}}$ \cite{Bartol96} & 
$R=$ Data/MC$_{\mbox{no~osc}}$\\[1ex]
\hline
{} &{} &{} &{}\\[-1.5ex]
Upthr. & 857 & 1169 & 0.73\\[1ex]
IU & 157 & 285 & 0.55\\[1ex]
ID+UGS & 262 & 375 & 0.70\\[1ex]
\hline
\end{tabular}
\caption{{\footnotesize  Summary of the MACRO $\nu _{\mu}$$\rightarrow$$ \mu$ events in
    $-1$$<$cos$\theta$$<$ 0 after background subtraction.
For each topology (see Fig. \ref{fig2}) the number of measured events, the MC
prediction for no-oscillations and the ratio (Data/MC-no osc) are given.}}
\end{center}
\end{table}

\begin{figure}
\begin{center}
  \mbox{ \epsfysize=7.2cm
         \epsffile{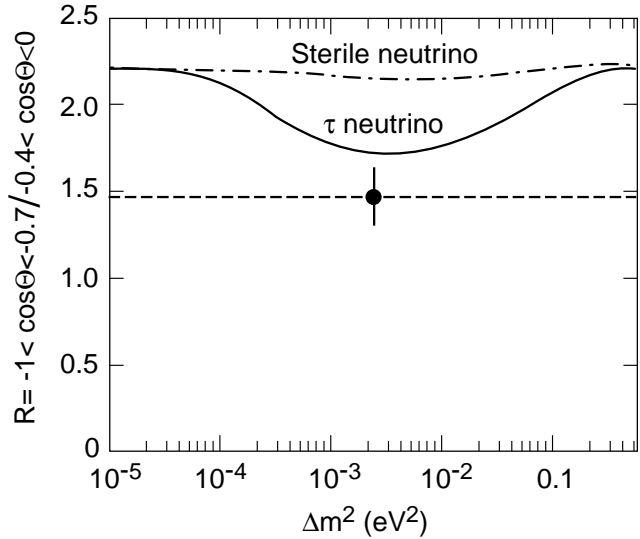} }
\caption{\footnotesize {Ratio of events with $-1< cos \theta < -0.7$ to events with
$-0.4<cos\theta<0$ as a function of $\Delta m^2$ for maximal mixing. The
black point with error bar is the measured value, the solid line is the
prediction for $ \nu_\mu \rightarrow \nu_\tau $ oscillations, the dash-dotted
line is the prediction for  $ \nu_\mu \rightarrow \nu_{sterile} $ oscillations.
\label{fig:fig8}}}
\end{center}
\end{figure}

\begin{figure}
\vspace{-2cm }
 \begin{center}
\mbox{\epsfig{figure=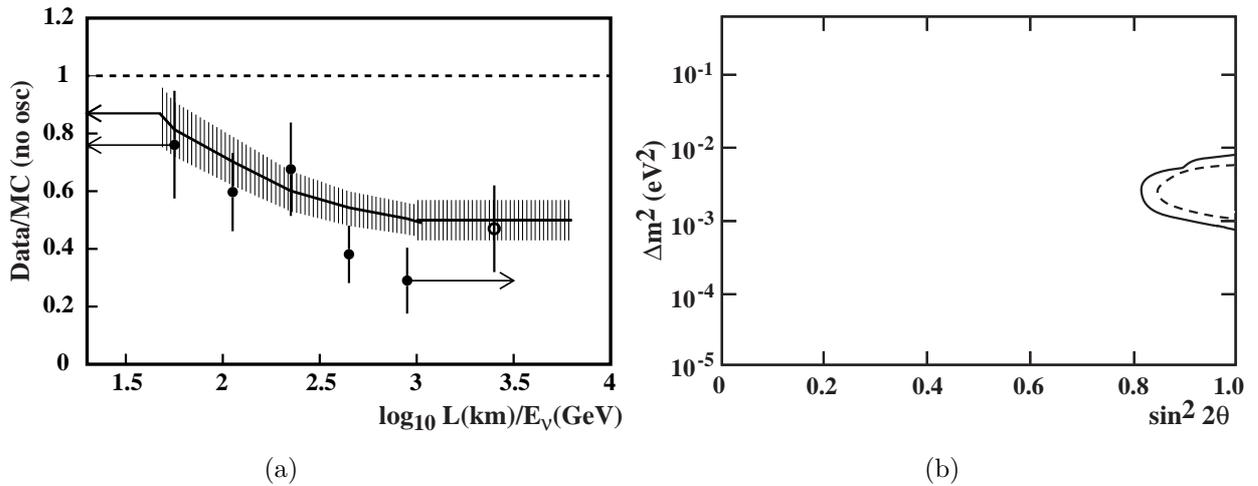,height=5.8cm}}
{\footnotesize  (a) \hspace {8cm} (b)}
\caption {\small(a) Data/MC$_{\mbox{no~osc}}$ 
vs $L/ E_\nu$ for  upthroughgoing muons (black points). The solid line is the MC expectation for 
$\Dm = 2.3 \cdot 10^{-3}$ eV$^2$ and $\stheta = 1$. The last point 
(empty circle) is obtained from the IU sample. (b) Interpolated 90\% c.l.
contour plots of the allowed regions in the $\Dm-\stheta  $ plane for the 
MACRO data using only the ratios $R_1,R_2,~R_3$ (continuous line) and adding
also the information on the absolute values $R_4,~R_5$ (dotted line).}
\label{fig:le-contour}
\end{center}
\end{figure}

Table 1 gives the measured and expected events for the 3  topologies. They all favor $\nu$ oscillations. The  $L/E_\nu$ distribution, Fig. \ref{fig:le-contour}a, deviates from MC expectations without oscillations; the deviations point to the same $\nmnt$  oscillation scenario \cite{mac31} \cite{mac34} \cite{mac52}.\\
{\bf \boldmath $\nmnt$ against $\nmns$.} Matter effects due to 
the difference between the weak interaction effective potential for 
muon neutrinos with respect to sterile neutrinos, which have null
potential, yield different total number and different zenith 
distributions of upthroughgoing muons \cite{mac34}. The measured 
ratio $R_{meas}$ between the events with $-1 < \cos \Theta < -0.7$ and with 
$-0.4 < \cos \Theta < 0$ was used \cite{mac34}. In this ratio most 
of the theoretical uncertainties on neutrino flux and cross sections cancel (combining the experimental
and theoretical errors in quadrature, a global uncertainty of 6\% is obtained).
 The measured ratio is $R_{meas}=1.38$, to be compared with $R_\tau=1.61$ and
$R_{sterile}=2.03$. One concludes that $\nmns$ oscillations (with any mixing) 
are excluded at the 99.8\% c.l. compared to the $\nmnt$ 
channel, see Fig. \ref{fig:fig8}.

{\bf \boldmath $\nm$ energy estimate by Multiple Coulomb Scattering (MCS) of upthroughgoing muons.}
Since MACRO was not equipped with a magnet, 
 the only way to estimate the muon energy was through their Multiple Coulomb 
Scattering (MCS) in the absorbers. Two analyses were 
performed \cite{mac41} \cite{mac48}. The first was made studying the deflection of 
muons using the streamer tubes in digital mode. This method 
had a spatial resolution of $\sim$1 cm. The second analysis was performed 
using the streamer tubes in ``drift mode". To check the electronics and the feasibility of the analysis two tests were performed at the CERN PS. The space resolution was $\simeq$3 mm. For 
each muon, 7 MCS variables were defined and given in input to a Neural 
Network,  trained with MC events of known energy 
crossing the detector at different zenith angles. The output of this program
gave the muon energy estimate event by event. The sample of upthroughgoing 
muons was separated in 4 subsamples with 
average energies $E_\mu$ of 12, 20, 50 and 100 GeV. The ratios 
Data/MC$_{\mbox{no~osc}}$ as a function of 
$\log_{10} (L/E_\nu)$ for upthroughgoing muons are plotted in 
Fig. \ref{fig:le-contour}a; they are in agreement with the 
$\nmnt$ oscillation hypothesis \cite{mac31} \cite{mac34} \cite{mac52}. 
\begin{figure}
  \begin{center}
 \vspace{-5cm }
   \mbox{\epsfysize=8cm    \epsffile{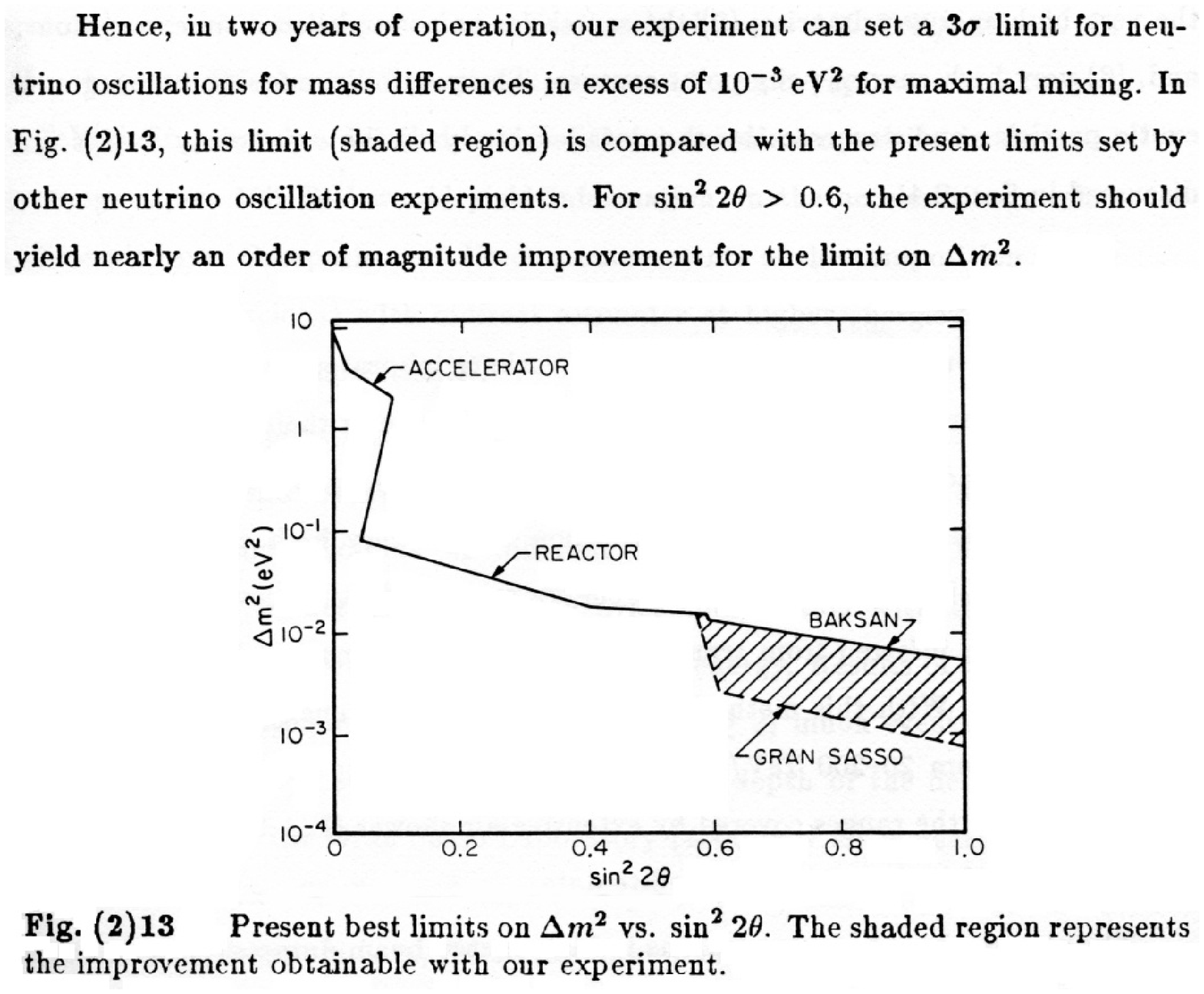}}
\caption{\footnotesize {From the 1984 MACRO proposal \cite{mac1}. \label{fig:fig_prop}}}
\end{center}
\end{figure}
{\bf New determination of the oscillation parameters.} In the early analyses MACRO
fitted the shape of the upthroughgoing muon zenith distribution and the 
absolute flux compared to Bartol96. This yielded $\Dm=2.5 \cdot 10^{-3}$ 
eV$^2$ and maximal mixing \cite{mac31} \cite{mac34}. Later, in order to reduce 
the effects of systematic uncertainties in the MC simulations, MACRO 
 used the following three independent ratios (it was checked
 that FLUKA, HKKM01 and Bartol96 MC yield the same 
predictions to within $\sim 5 \%$):

(i) High Energy Data: zenith distribution ratio: $R_1 = N_{vert}/N_{hor}$

(ii) High Energy Data: $\nu$ energy measurement ratio: $R_2 = N_{low}/N_{high}$

(iii) Low Energy Data:  $R_3 = (Data/MC)_{IU}/(Data/MC)_{ID+UGS}$.

\noindent The no oscillation hypothesis has a probability 
P$\sim$3 $\cdot 10^{-7} $
and is thus ruled out by $ \sim 5 \sigma$. By fitting the 3 ratios 
to the $\nmnt$ oscillation formulae, we obtained $\stheta = 1,~\Dm = 2.3
\cdot 10^{-3}$ eV$^2$ and the allowed region indicated by the solid line 
in Fig. \ref{fig:le-contour}b. There is  good consistency between the old and 
new methods \cite{mac52}. 
Using  Bartol96, it is possible to add the 
information on absolute fluxes:

   (iv) High energy data (systematic error $\simeq$17$\%$): $R_4 = N_{meas}/N_{MC}$.

   (v) Low energy semicontained muons (scale error $21 \%$): $R_5 = N_{meas}/N_{MC}$.

\noindent These informations leave the best fit values unchanged and reduce the area of the allowed region, as indicated by the dashed line in Fig. \ref{fig:le-contour}b. 
($6\sigma$ significance). \\

Fig. \ref{fig:fig_prop} shows one page of the 1984  proposal \cite{mac1}. It  stated that MACRO could yield valuable information on possible neutrino oscillations in the then unexplored region indicated by the shaded area: in fact atmospheric neutrino oscillations were found  there! \cite{gg}.

{\bf Exotic oscillations.} MACRO and SuperK data have been used to search for sub-dominant oscillations due to possible Lorentz invariance violation (or violation of the equivalence principle). In the first case there could be mixing between flavor and velocity eigenstates. The data disfavor these exotic possibilities, even at a sub-dominant level. Stringent 90$\%$ c.l. limits were placed in the Lorentz invariance violation parameters 
$\left|\Delta v\right|<6\cdot10^{-24}$ at ${\stheta}_v$=0 and $\left|\Delta v\right|<4\cdot10^{-26}$ at ${\stheta}_v=\pm$1 \cite{mac38} \cite{gonzalez}.

{\bf Neutrino decay.} It could be another exotic partial explanation for neutrino disappearence; no radiative decay has been observed \cite{cecchini}.

\section{Search for Astrophysical Point Sources of HE \( \nu _{\mu } \)'s}

\begin{figure}
\begin{center}
\vspace{-2cm }
\hspace{-3cm }
  \mbox{ 
\epsfysize=6.6cm
         \epsffile{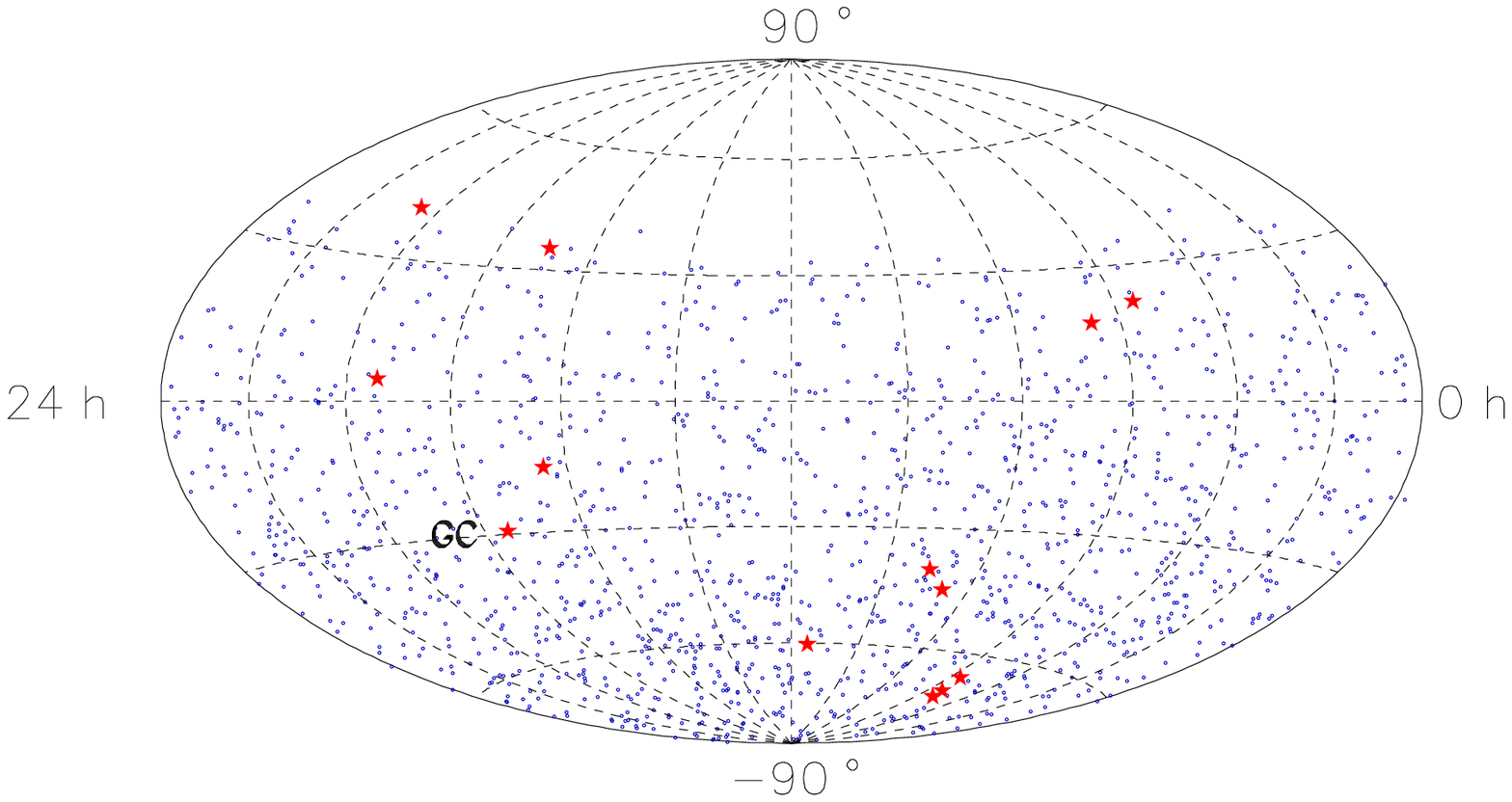}
\hspace{-2.1cm }
\epsfysize=8cm
         \epsffile{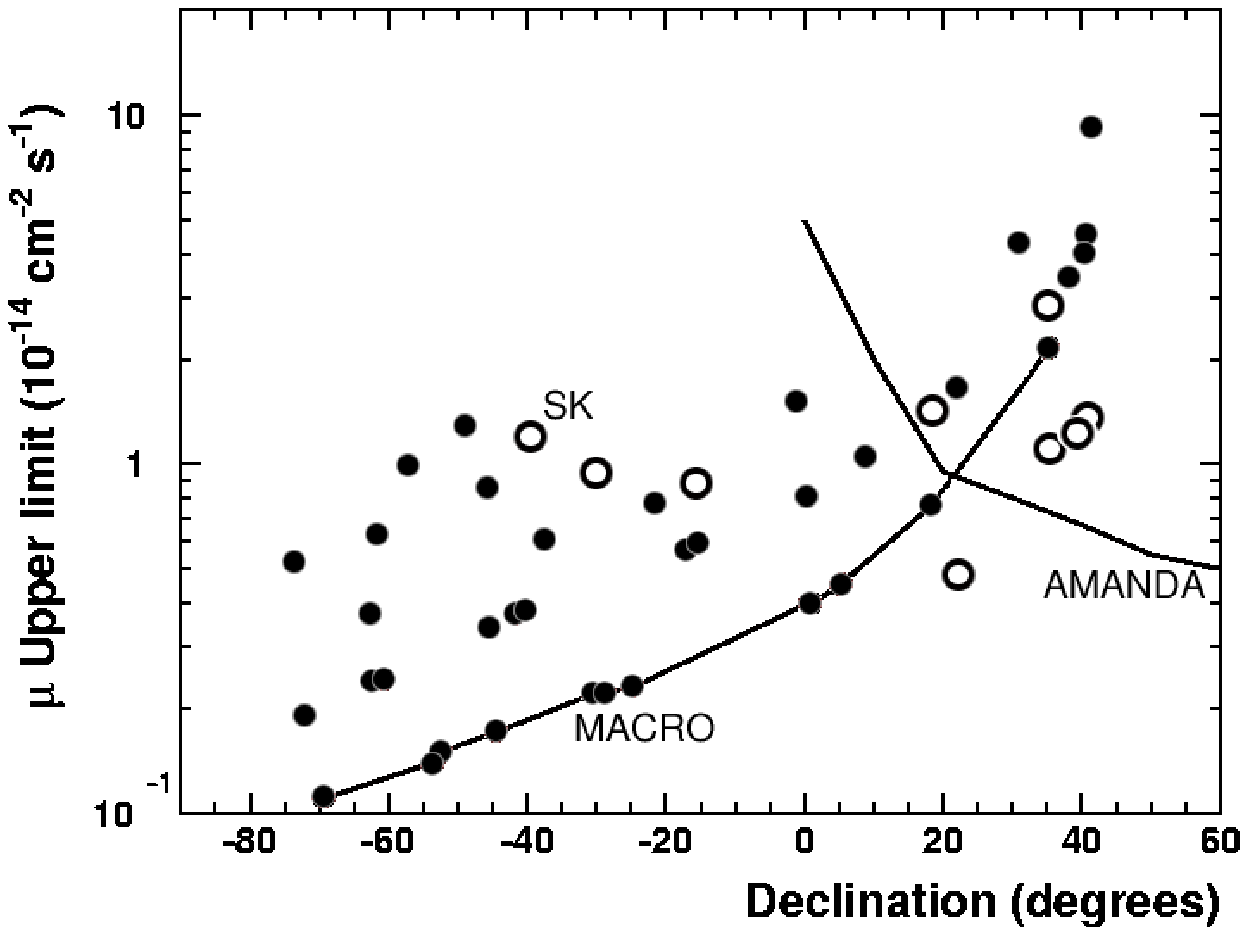} }
{\small \hskip 8.0 truecm (a) \hskip 7.5 truecm (b)}
\caption{\label{fig11}\footnotesize {
High energy neutrino astronomy. (a) Upgoing muon distribution in equatorial
coordinates (1356 events). 
(b) The black points are the MACRO 90 \% c.l. upgoing 
muon flux limits vs declination for 42 selected  point sources. The solid line refers to the
limits obtained for those cases for which the atmospheric neutrino
background was zero. The limits from the SK (open circles) and AMANDA (thin
line) experiments are quoted; these  limits  refer to  higher $E_\nu$.
}}
\end{center}
\end{figure}

High energy \( \nu _{\mu }\)'s are expected to come from several
galactic and extragalactic sources. Neutrino production requires
astrophysical accelerators of charged particles and astrophysical beam
dumps. The good angular resolution of MACRO allowed a
sensitive search for upgoing muons produced by neutrinos coming from
celestial sources, with a negligible atmospheric neutrino background.
An excess of events was searched for around the positions of known
sources in \( 3^{\circ } \) (half width) angular bins. This value
was chosen to take into account the angular smearing produced
by the multiple muon scattering in the rock below the detector and by the
energy-integrated angular distribution of the scattered muon with
respect to the neutrino direction. In a total livetime of 6.16 y  we obtained 1356 events,
 Fig. \ref{fig11}. The 90\% c.l. upper limits on the muon fluxes from
specific celestial sources lay in the range \( 10^{-15}-10^{-14}{\cm }^{-2}{\s }^{-1} \), 
Fig. \ref{fig11}b.
The solid  line is MACRO sensitivity vs. declination. Notice the
two cases, GX339-4 ($\alpha = 255.71^o $, $\delta= -48.79^o $)  and Cir
X-1  ($\alpha = 230.17^o $, $\delta= -57.17^o $), with 7 events: in Fig. \ref{fig11} they are
considered as  background, thus the upper flux limits
are higher; but they could also be indications of signals \cite{mac33}.

We searched for time coincidences of our upgoing muons with \( \gamma  \)-ray
bursts as given in the BATSE 3B and 4B catalogues, for the period
from 1991-2000 \cite{mac33}. No statistically significant
time correlation was found.

We have also searched for a diffuse astrophysical neutrino flux for
which we establish a flux upper limit at the level of \( 1.5\cdot
10^{-14}{\cm }^{-2}{\s }^{-1} \) \cite{mac39}.

\section{Indirect Searches for WIMPs}
Weakly Interacting Massive Particles (WIMPs) may be part of the
galactic dark matter; they may be intercepted by celestial bodies,
slowed down and trapped in their centers, where WIMP and anti-WIMP could
annihilate  yielding upthroughgoing muons. The annihilations  yield $\nu_{\mu}$ of \GeV{} energy,
in small angular windows from their centers. 

\begin{figure}
 \begin{center}
  \mbox{ \epsfysize=6cm
         \epsffile{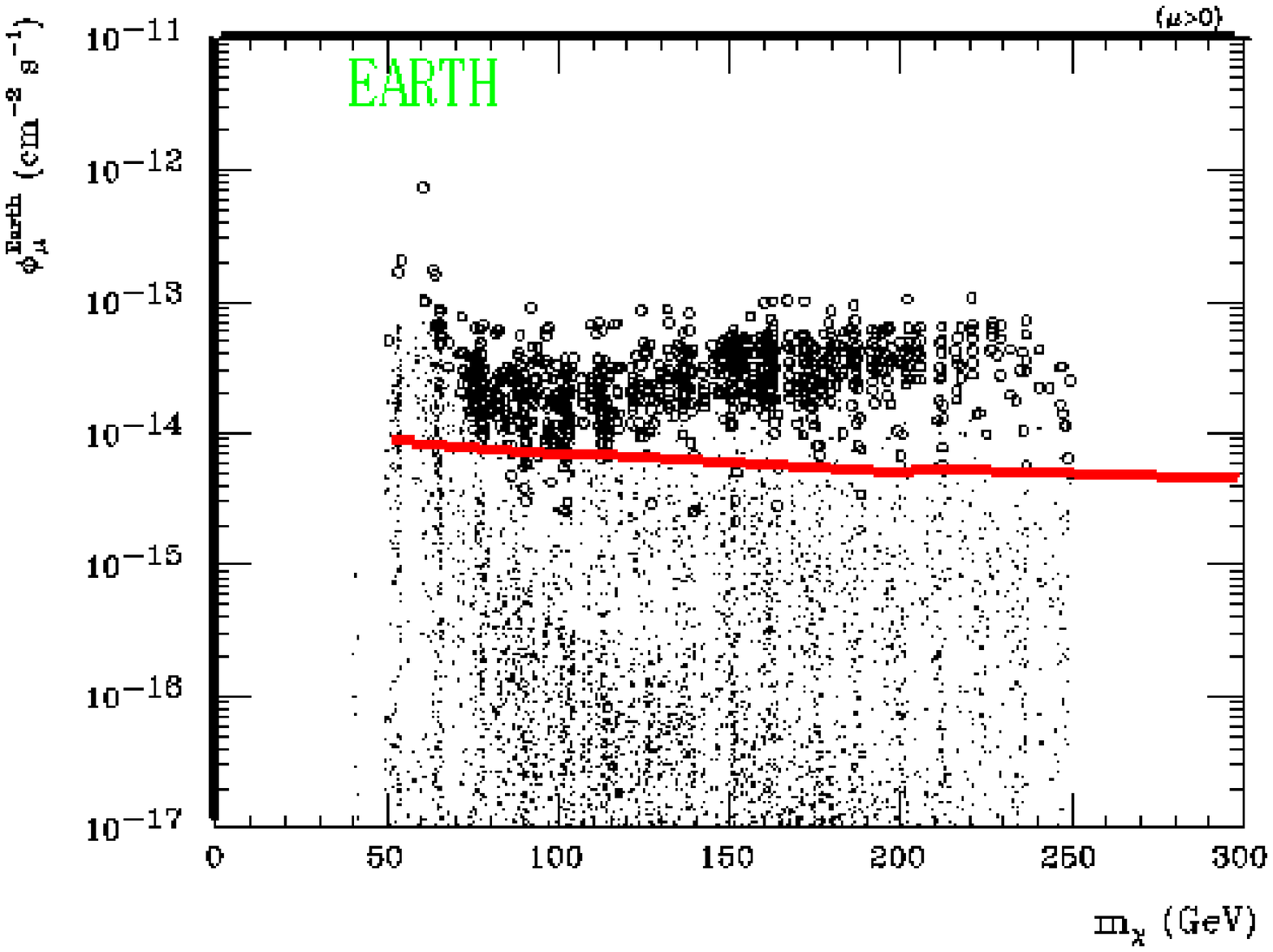}
\epsfysize=6cm
         \epsffile{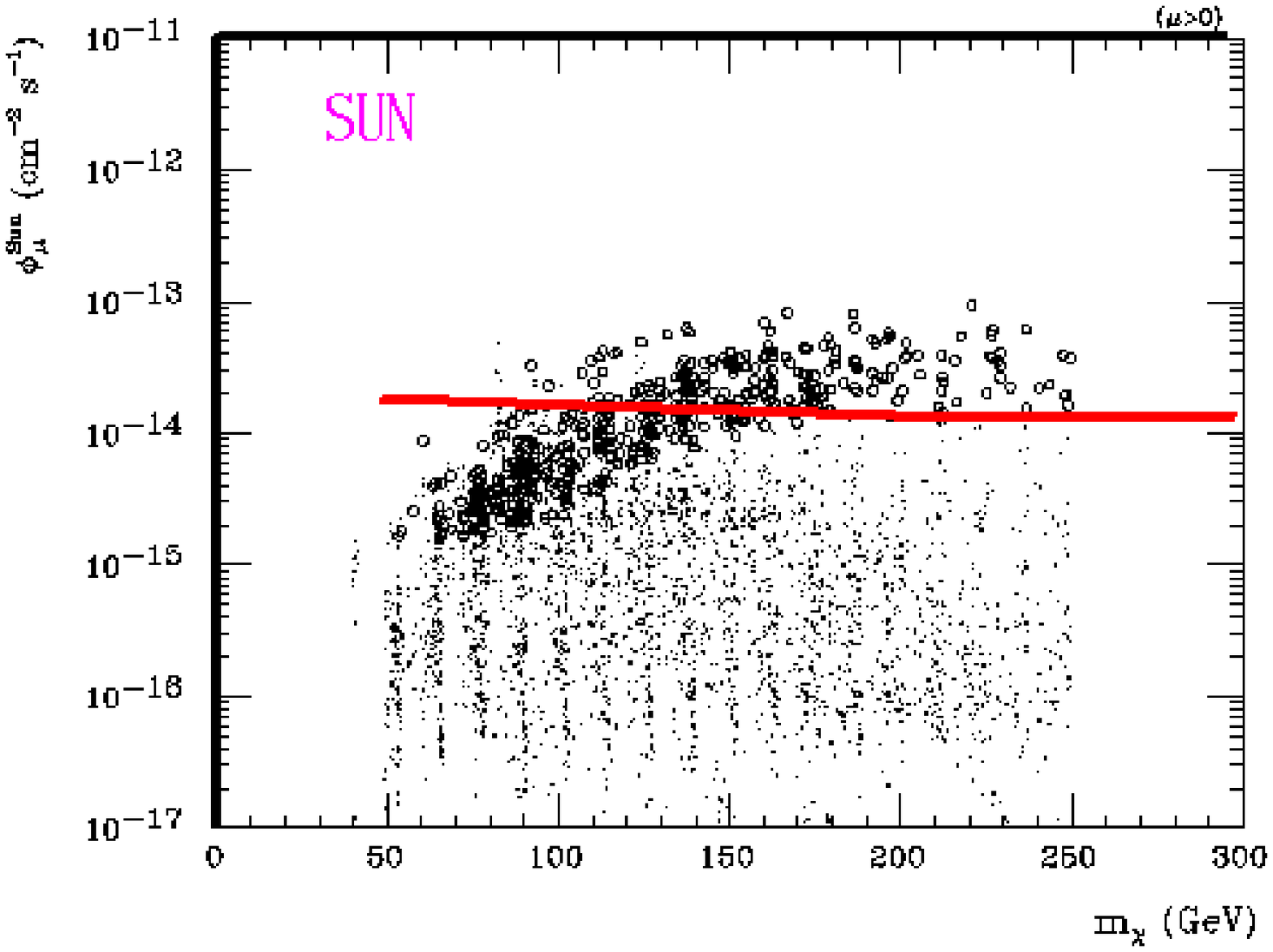}   }
 \end{center}
{\centering 
{\small \hskip 5.0 truecm (a) \hskip 7.5 truecm (b)}}
\caption{\label{fig12}\footnotesize  {
(a) The solid line is our upwardgoing muon flux upper limit (90\% c.l.) from the Earth plotted
vs. neutralino mass $m_\chi$  (${E_\mu }^{th}$=1$\GeV$).  (b) As
in (a) but for upwardgoing muons from the Sun \cite{mac28}.
Each dot is obtained varying model parameters. The open circles indicate
models {\it excluded  } by direct measurements, assuming a local dark matter density of $\sim  0.5 \GeV cm^{-3}$.}}
\end{figure}

For the Earth we have chosen a \( 15^{o} \) cone around the vertical:
we found 863 events. The MC expectation for atmospheric \( \nu _{\mu } \)
without oscillations gave a larger number of events. We set a conservative
flux upper limit assuming that the measured number of events equals the
expected ones. The \( 90\, \% \) c.l. limits for
the flux of upgoing muons are shown in Fig.  \ref{fig12}a (it varies from about 0.8
to 0.5 \( 10^{-14}{\cm }^{-2}{\s }^{-1} \)). If the WIMPs are identified
with the smallest mass neutralino, the MACRO limit may be used to
constrain the stable neutralino mass, following the model of Bottino
et al. \cite{bottino} \cite{mac28}, see Fig. \ref{fig12}a.

A similar procedure was used to search for \( \nu _{\mu } \) from the
Sun, using 10 search cones. In the absence of statistically significant excesses muon upper limits
at the level of  \( 1.5-2\cdot 10^{-14}{\cm }^{-2}{\s }^{-1} \) were
established, see Fig.  \ref{fig12}b.

\section{Magnetic Monopoles and Nuclearites}

The concept of magnetic monopole (MM) may be traced back to the origin of magnetism. In 1931 Dirac introduced MMs in order to explain the quantization of the electric charge, obtaining the formula e$g=n\hbar c/2$, from which 
$g=ng_D=n\hbar c/2 = n 3.29 \cdot 10^{-8}$ c.g.s.; n is an integer \cite{dirac}.  

Supermassive monopoles predicted by Grand Unified Theories (GUT) of the electroweak and strong interactions should have 
masses of the order of \( m_M\)$\sim$$10^{17}{\GeV } $ \cite{thooft}. These MMs could be present in the penetrating cosmic radiation and are expected to arrive isotropically from all directions and to have typical galactic velocities, $\sim$10$^{-3}c$, if trapped in our Galaxy. MMs trapped in our solar system or in the supercluster of galaxies
may travel with typical velocities of the order of $ \sim$$ 10^{-4}c$
and $\sim$$ 10^{-2}c$, respectively \cite{giapat}. 

The search for GUT magnetic monopoles  was one of the main objectives of MACRO. 

Monopoles, in the presence of strong magnetic fields, may reach higher velocities. Possible Intermediate
Mass MMs could achieve relativistic velocities, but they could reach MACRO only from above.

The reference sensitivity level for a significant MM search is the
Parker bound \cite{giapat}, the maximum monopole flux compatible with the survival
of the galactic magnetic field. This limit is of the order of \( \Phi \lsim 10^{-15}{\cm }^{-2}{\s }^{-1}{\sr }^{-1} \),
but it could be reduced by almost an order of magnitude when considering
the survival of a small galactic magnetic field seed.
Our experiment was designed to reach a flux sensitivity for GUT MMs well below
the Parker bound, in the MM velocity range \( 4\times 10^{-5}<\beta <1. \)
The three MACRO sub-detectors had sensitivities in wide \( \beta  \)-ranges,
with overlapping regions; thus they allowed multiple signatures of the
same rare event candidate. 

\begin{figure}
\vspace{-1.5cm}
 \begin{center}
  \mbox{ \epsfysize=10cm \epsffile{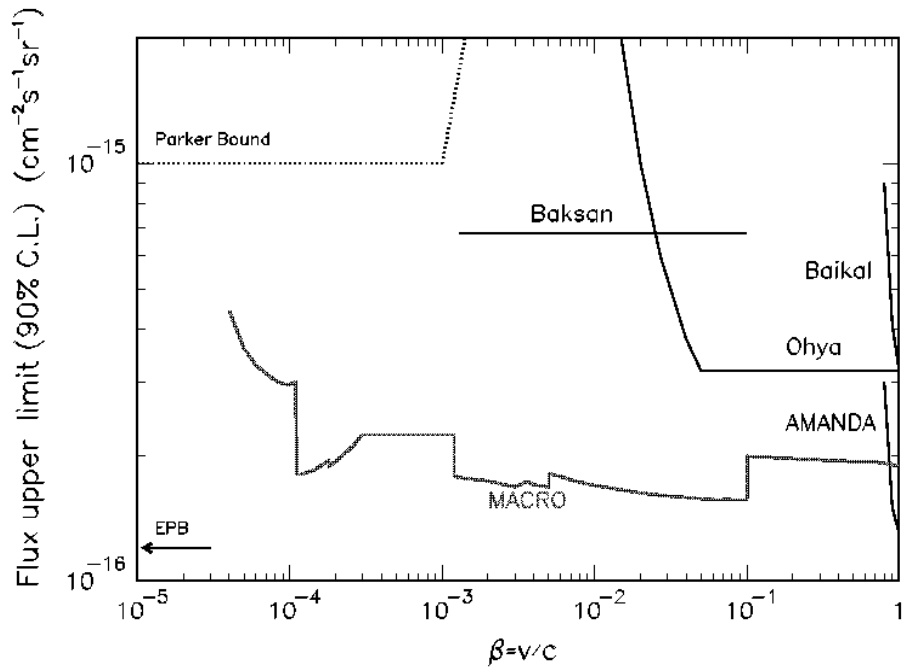}}
 \end{center}
\caption{\label{fig13}\footnotesize  {
(a) Magnetic monopole flux upper limits at the $90\%$ c.l. obtained by MACRO and by other experiments. The limits apply to singly charged ($ { g =g_D }$) monopoles assuming that catalysis cross sections are smaller than a few mb. (b) Limits obtained with the MACRO NTDs only, for GUT monopoles with $g=g_D, 2g_D,3g_D$ and for dyons.}}
\end{figure}

No candidate was found in many years
of data taking by any of the three subdetectors. The MM flux limits set by different analyses using the three subdetectors over different \( \beta  \)-range were combined to obtain a global
MACRO limit. For each \( \beta  \) value, the global time integrated
acceptance was computed as the sum of the independent portions of
each analysis. The limits versus \( \beta  \) are shown in Fig. \ref{fig13} 
together with the limits set by other experiments \cite{giapat, nucl1, nucl2}; other limits 
are quoted in \cite{mac42}. The limits obtained with the MACRO NTDs only are shown in Fig. \ref{fig13}b:
our  limits are the best direct limits existing for GUT MMs over a wide range of \( \beta  \) , \( 4\times 10^{-5}<\beta <1; \) see ref. \cite{nucl3} \cite{milton} for stringent limits with indirect experiments. Energy losses are computed in ref. \cite{gg} \cite{derka}.

A specific search for GUT monopole catalysis of nucleon decay was made with our
streamer tube system. Since no event was found, one can place a
monopole flux upper limit at the level of $\sim 3 \cdot 10^{-16}
\,cm^{-2}s^{-1}sr^{-1}$ for $10^{-4} \lsim \beta \lsim 5 \cdot 10^{-3}$,
valid for a large catalysis cross section, $5\cdot 10^2 < \sigma_{cat}
< 10^3\, mb$ \cite{mac43}. The flux limit for the  standard direct MM search with
streamer tubes is valid for  $\sigma_{cat} < 100\,mb$.

The MM searches based on the scintillator and on the nuclear track subdetectors
were also used to set new upper limits on the flux of cosmic ray \textit{nuclearites}
(strange quark matter), over the same \( \beta  \) range.
If nuclearites are part of the dark matter in our galaxy, the most
interesting \( \beta  \) is of the order of \( \sim 10^{-3} \)\cite{witten}.  
Some of the nuclearite limits apply also to charged \textit{Q-balls} (agglomerates
of squarks, sleptons and Higgs fields and other objects)\cite{coleman}\cite{klopov}\cite{corint}. 

The energy losses of MMs, dyons and of other heavy particles in the 
Earth and in different detectors for various particle masses and 
velocities were computed in \cite{derka}.

\section{Neutrinos from Stellar Gravitational Collapses}

A stellar gravitational collapse (GC) of the core of a massive star
is expected to produce a  burst of all types of ${\nu}$ and $\bar{\nu}$
with energies of \( 7-30 \)~\MeV{} and  a duration
of 10 s. The \anue{}'s can be detected via the process \( \bar{\nu }_{e}+p\rightarrow n+e^{+} \) in  liquid scintillators; \( 100\div 150 \) \anue{} events were expected in our
 scintillator for a stellar collapse at the center of our Galaxy.

We used two electronic systems for detecting \anue{}'s from stellar
gravitational collapses. The first system was based on the dedicated
PHRASE trigger, the second one was based on the ERP trigger, see sect. 2. Both
systems had an energy threshold of $\sim$7$\MeV$ and recorded
pulse shape, charge and timing informations. Immediately after a $ >$7$\MeV$
trigger, the PHRASE system lowered its threshold to about 1 MeV, for
 \( 800{\mu \s } \),  to detect (with a \( \simeq 25\, \% \)
efficiency) the \( 2.2{\MeV } \) \( \gamma  \) released in the reaction
\( n\, +\, p\rightarrow d\, +\, \gamma _{2.2\MeV } \) induced by
the neutron produced in the primary process.

A redundant supernova alarm system was in operation, alerting immediately
the physicists on shift. We defined a general procedure to alert the
physics and astrophysics communities in case of an interesting alarm. Finally, a procedure to link the various supernova
observatories around the world was set up \cite{mac23}.

The effective MACRO active mass was $\sim$580 t; the live-time fraction
in the last four years was $\simeq$97.5 \%. No stellar gravitational
collapse was observed in our Galaxy from the beginning of \( 1989 \)
to the end of 2000 \cite{mac47}.

\section{Cosmic Ray Muons}
MACRO large area and acceptance allowed to record \( 6\cdot 10^{7} \) single muons, \( 3.7 \cdot 10^{6} \) multiple muons and to study 
many aspects of physics and astrophysics of cosmic rays (CR). \\
\noindent \textbf{Muon vertical intensity.} The underground muon vertical
intensity vs rock thickness gives information on the high energy
(\( E\, \, \gsim \, \, 1.3{\TeV } \)) atmospheric muon flux and on
the all-particle primary CR spectrum. The results  constrain
the CR production and interaction models. A study
performed in \( 1995 \) covered the overburden range \( 2200\div 7000\hg /\cm ^{2} \)
\cite{mac16}.

\begin{figure}
\vspace{-1cm}
  \begin{center}
  \mbox{ 
\hspace{-1cm}
 \epsfysize=9cm
         \epsffile{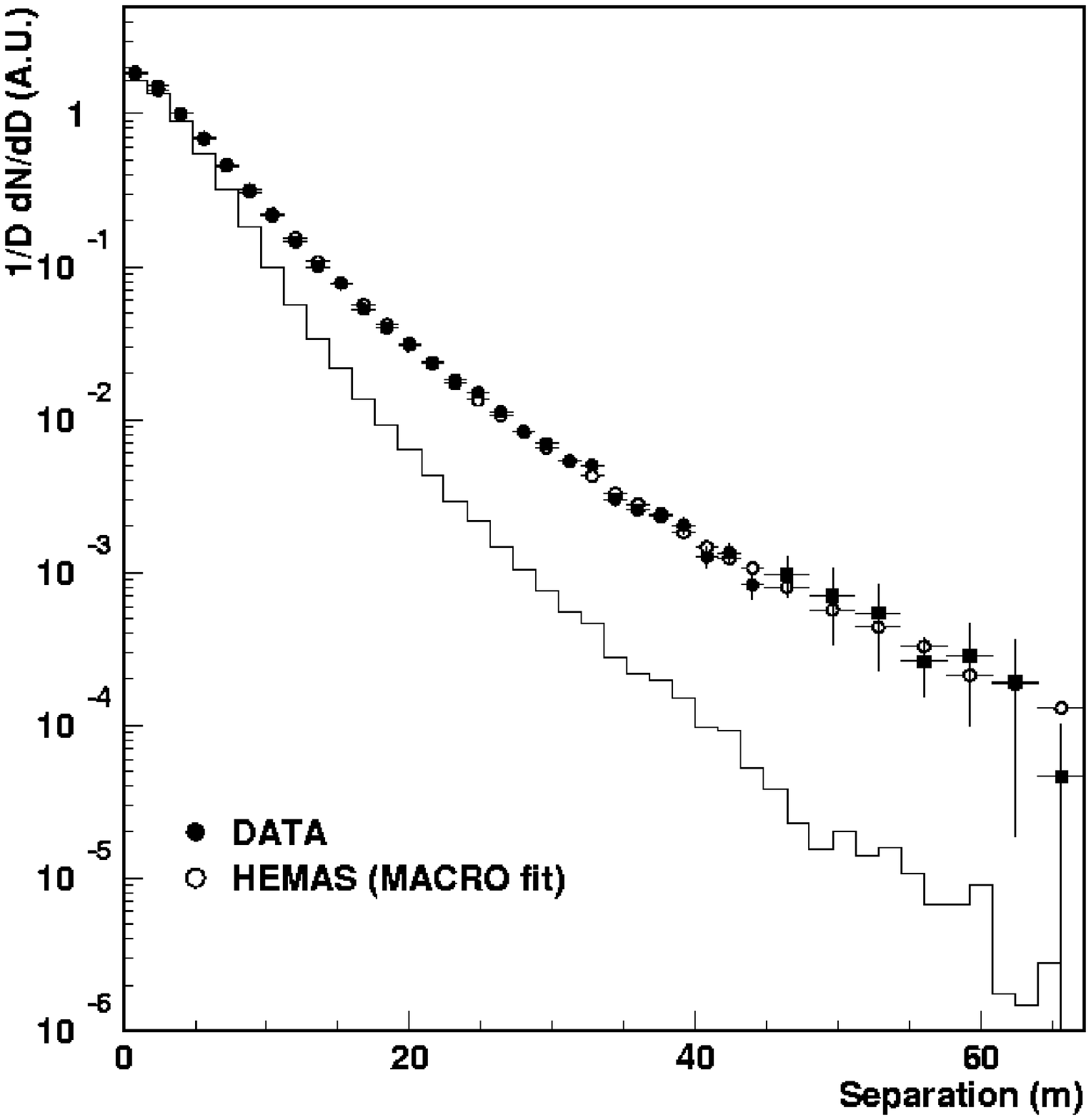} 
 \epsfysize=9cm
         \epsffile{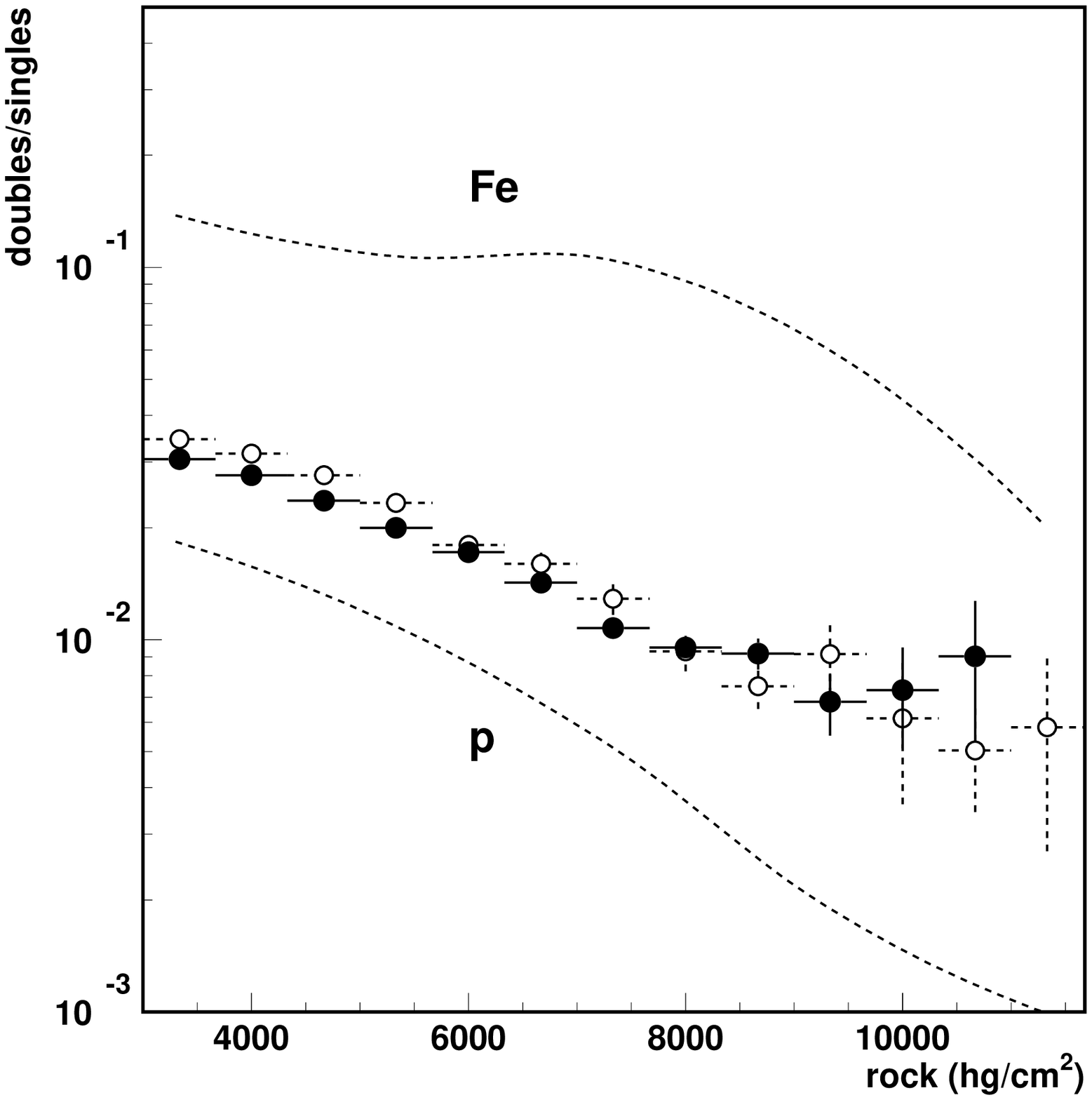} 
}
{\small \hskip 4.0 truecm (a) \hskip 7.5 truecm (b)}   
\caption{\label{fig15}\footnotesize {(a) Unfolded experimental decoherence 
distribution for an infinite detector (black points) compared with MC 
expectations (open points); the histogram is the measured decoherence distribution before 
unfolding  \cite{mac29}.
(b) Ratio of double  to single muon events vs the rock depth. The black points are our data; the open circles are
MC predictions made using the MACRO composition model. The dashed lines are MC predictions using
pure proton and iron primaries.}}
\end{center}
\end{figure}

\noindent \textbf{Analysis of high multiplicity muon bundles.} The
 \textbf{multiplicity distribution} of muon bundles provides
information on the primary CR composition model. 
The  \textbf{decoherence} function (the distribution of the distance
between two muons in a muon bundle) gives  information on the hadronic interaction
features at high energies; a study was performed using a large sample
of data and improved MC methods,  Fig. \ref{fig15}a \cite{mac29}.
Different hadronic interaction 
models (DPMJET, QGSJET, SIBYLL,
HEMAS, HDPM) interfaced to the HEMAS and CORSIKA shower propagation
codes were used.

\emph{Muon correlations inside a bundle}  were studied, 
using the so called correlation integral 
\cite{corint}, to search for dynamical correlations in the
bundles.  Since the cascade development in atmosphere is mainly determined by the number of {}``steps{}''
in the {}``tree formation{}'', we expect a different behaviour for cascades originated by  light and
heavy CR primaries. For the same reason, the analysis
is less sensitive to the hadronic interaction model used
in the simulations. For $E_{pr} >  1000$ TeV, the composition model derived from the
analysis of the muon multiplicity distribution \cite{mac19, mac20}
is almost independent of the interaction model.

\par We also  searched for  \emph{substructures ({}``clusters{}'')
inside muon bundles} \cite{clusters}. The search  
was performed by means of different software algorithms; the study
is sensitive to both  hadronic interaction  and   primary CR
composition models. 
If the primary composition has been determined by the first method, a choice of the bundle topology
gives interesting connections with the early hadronic interaction
features in the atmosphere.  The comparison between our data and MC simulations allowed to place constraints on  the  interaction models. The same MC study has
shown that muon bundles with a central core and an isolated cluster
with at least two muons are the result of random associations. An analysis  of the  decoherence
function for high multiplicity events has shown that QGSJET is the hadronic
interaction model which better reproduces the underground observables.\\
\textbf{The ratio double muons/single muons:} The ratio $N_2$/$N_1$ of double muon  over
single muon events is expected to decrease at increasing  rock
depths.  
The LVD collaboration reported an increase of the ratio  multiple-muons to all-muons   for rock depths $h> 7000~hg/cm^2$.
We measured the ratio  as a function of the rock depth, using also multiple muon events at
large zenith angles. A detailed MC simulation was made using the HEMAS code, with the  zenith
angle  extended up to $89^\circ$.
The event direction
was reconstructed by the tracking system. The
rock depth is provided by the Gran Sasso map function $h(\theta,\phi)$ up to $\theta =94^\circ$.   MC simulations have shown that the percentage
of events with mis-reconstructed multiplicity is $< 3\%$. Attention was 
devoted to  ``cleaning'' the events from spurious effects;  we also made a
visual scanning of the events.
Our measured ratios $N_{2}$/$N_{1}$ as a function of the rock depth,
Fig. \ref{fig15}(b), are 
in agreement with the expectation of a monotonic decrease
 up to $\sim 10000~hg/cm^{2}$. Above this value,
the low statistics does not allow  a firm conclusion
on a possible increase of $N_{2}$/$N_{1}$ \cite{clusters}.\\
\textbf{Muon Astronomy \cite{mac38}.}   Some past experiments reported 
excesses of a modulated muon flux from the direction of  Cyg X-\( 3 \). Our data did not indicate
 excesses above  background, both for steady
dc fluxes and for modulated ac fluxes.
\begin{figure}
  \begin{center}
  \mbox{ 
\epsfysize=8cm
         \epsffile{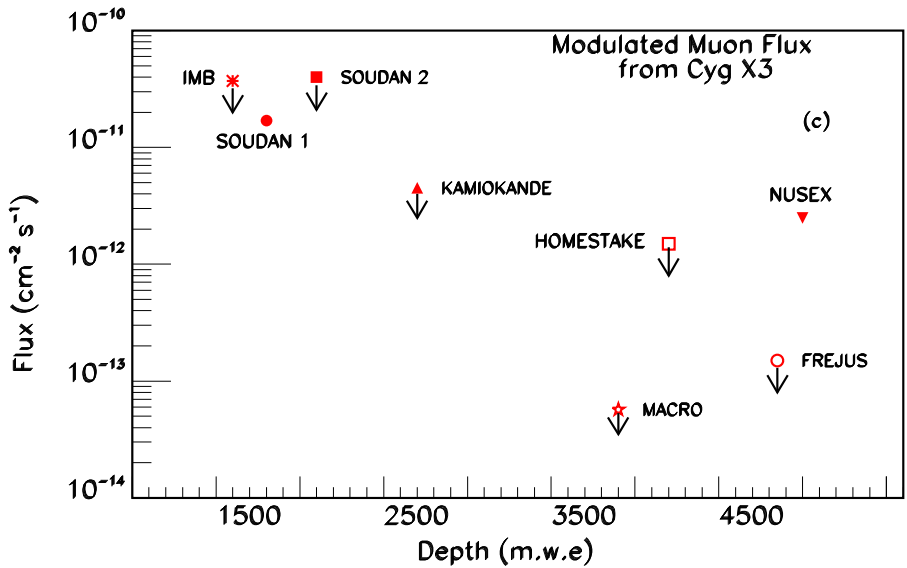} 
}   
\caption{\label{fig16}\footnotesize {Searches for a
  modulated muon signal from Cyg X-3. The Soudan 1 and Nusex collaborations
  reported positive indications, while  other experiments give flux
  upper limits.}}

\end{center}
\end{figure}
Our pointing precision was checked via the shadow of the Moon
and of the Sun on primary CRs.  The pointing resolution
was checked with double muons. The  angle
containing  \( 68\% \) of the events in a \( \Delta \theta  \) bin
was \( 0.8^{o} \).

\begin{figure}
  \begin{center}
  \mbox{ 
\epsfysize=10cm
         \epsffile{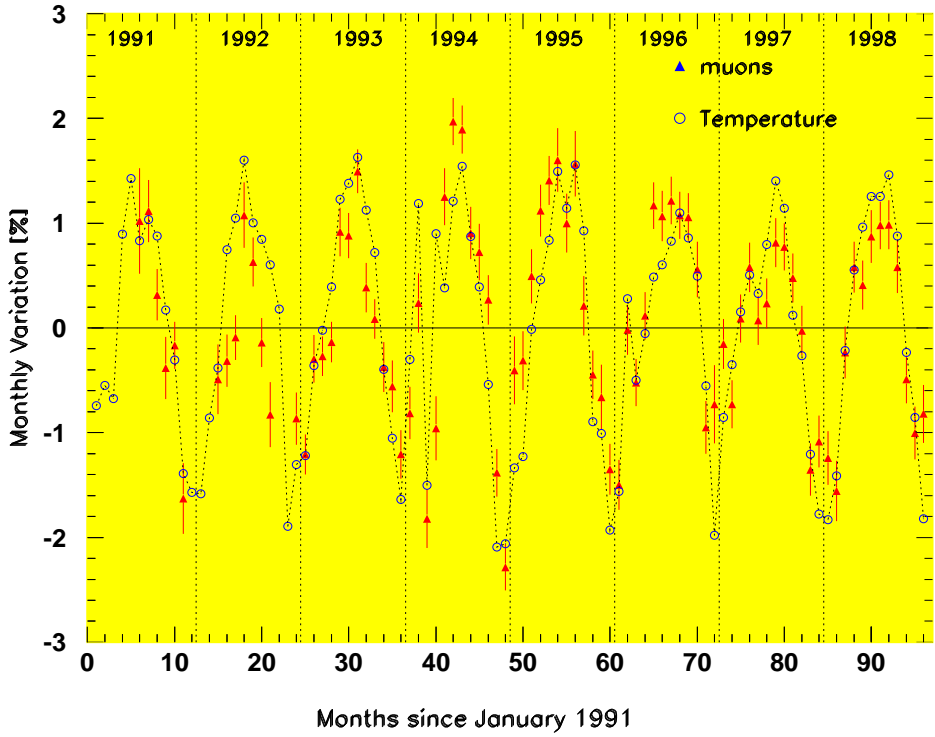} 
}   
\caption{\label{fig17}\footnotesize {
Seasonal variation of the muon flux from above (black triangles); the open circles are measurements of the temperature of the upper atmosphere.}}
\end{center}
\end{figure}

\begin{figure}
 \begin{center}
  \mbox{ \epsfysize=8cm
         \epsffile{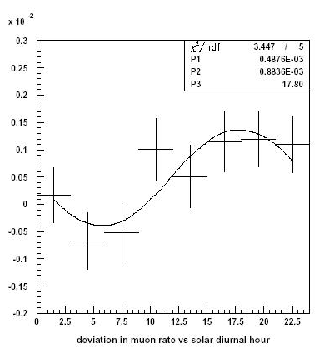}   
\epsfysize=8cm
         \epsffile{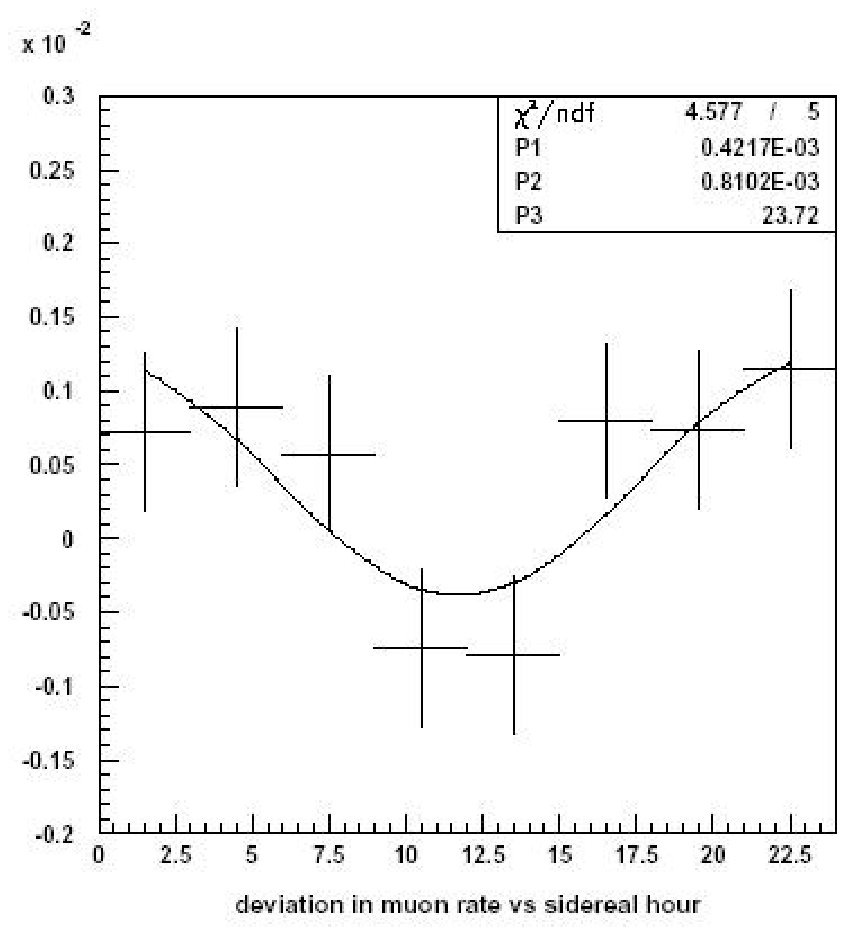}   }
 \end{center}
{\centering 
{\small \hskip 4.0 truecm (a) \hskip 7.5 truecm (b)}}
\caption{\label{fig18}\footnotesize {  
Deviations of the muon rate from the mean muon rate (a) versus the local
solar diurnal time at  Gran Sasso, and (b) versus the local sidereal time.
}}
\end{figure}

\noindent \textit{All sky d.c. survey.} The sky, in galactic coordinates,
was divided into bins of equal solid angle,
\( \Delta \alpha =3^{o} \), \( \Delta sin\delta =0.04 \),  corresponding
to  cones of \( 1.5^{o} \) half angles. In order to remove
edge effects, three other surveys were done, by shifting the map by
one-half-bin in \( \alpha  \), by one-half bin in sin\( \delta  \)
 and with both \( \alpha  \) and sin\( \delta  \) shifted. For each solid angle bin we computed the deviation from the
average  muon intensity, after background subtraction. 
No deviation was found; the $95\%$ flux upper limits were 
$\leq 5\times 10^{-13}cm^{-2}s^{-1}$.\\
\textit{Specific point-like d.c. sources}. For 
Cyg X-3, Mrk421, Mrk501 we searched in a narrow cone (\( 1^{o} \)
half angle) around the source direction. We obtained flux limits at
the level of \( (2-4)\cdot 10^{-13}cm^{-2}s^{-1} \). There is a
small excess of \( 2.0\, \, \sigma  \) in the direction
of Mrk501.\\
\textit{Modulated a.c. search from Cyg X-3 and Her X-1}. No evidence
for an excess was observed and the limits are \( {\Phi <2\times 10^{-13}cm^{-2}s^{-1}} \),
see Fig. \ref{fig16} .\\
\textit{Search for bursting episodes}.   A search was made for pulsed muon signals in a 
\( 1^{o} \) half angle cone around the location of high energy photon sources. Bursting episodes of duration of
\( \sim\)1 day were searched for with two different methods. In the
first  we searched for daily excesses of muons above the background,
also plotting cumulative excesses day by day. In the second method
we computed day by day the quantity \( -Log_{10}P \) where \textit{P}
is the probability to observe a burst at least as large as \( N_{obs} \).
We find some possible excesses for Mrk421 on the days 7/1/93, 14/2/95,
27/8/97, 5/12/98.  \\
 \textit{Seasonal variations.} Underground muons are produced by mesons
decaying in flight in the  atmosphere. The muon flux thus depends
on the ratio between  decay and  interaction probabilities of
the parent mesons, which are sensitive to the atmospheric density
and  temperature. The flux  decreases in
winter, when the temperature is lower and the atmosphere more dense,
and  increases in summer; the  variations are of \( \pm 2\% \), Fig. \ref{fig17}.\\
\textit{Solar daily variations.} Because of changes in the day-night
temperatures we expect solar daily variations similar to seasonal
variations, but of  smaller amplitudes. These variations were found with an amplitude $A = (0.88 \pm 0.26) \cdot 10^{-3}$ at a significance of  3.4 $\sigma$,  Fig.  \ref{fig18}a.
\begin{figure}
 \begin{center}
  \mbox{ \epsfysize=8cm
         \epsffile{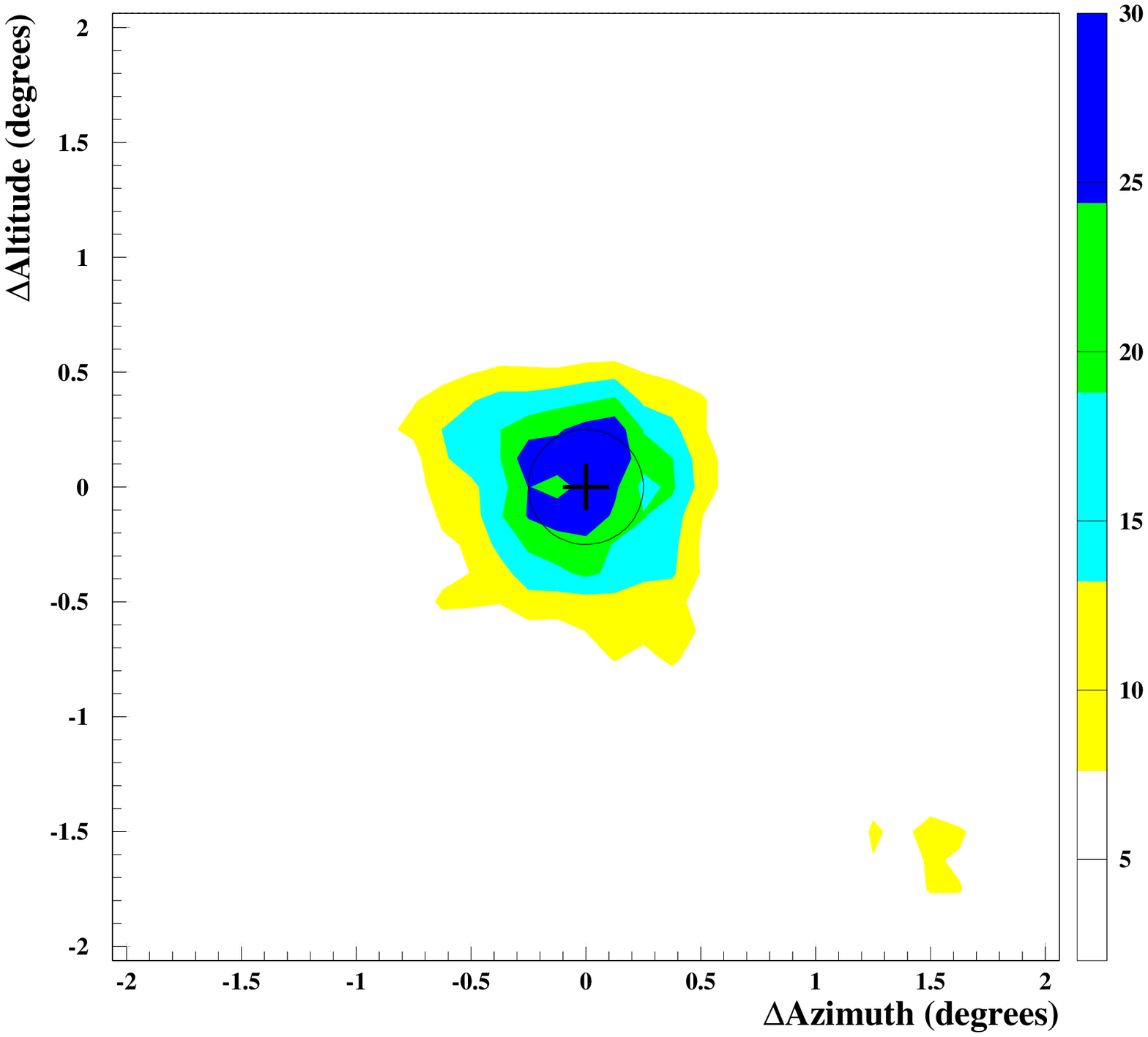}
\epsfysize=8cm
         \epsffile{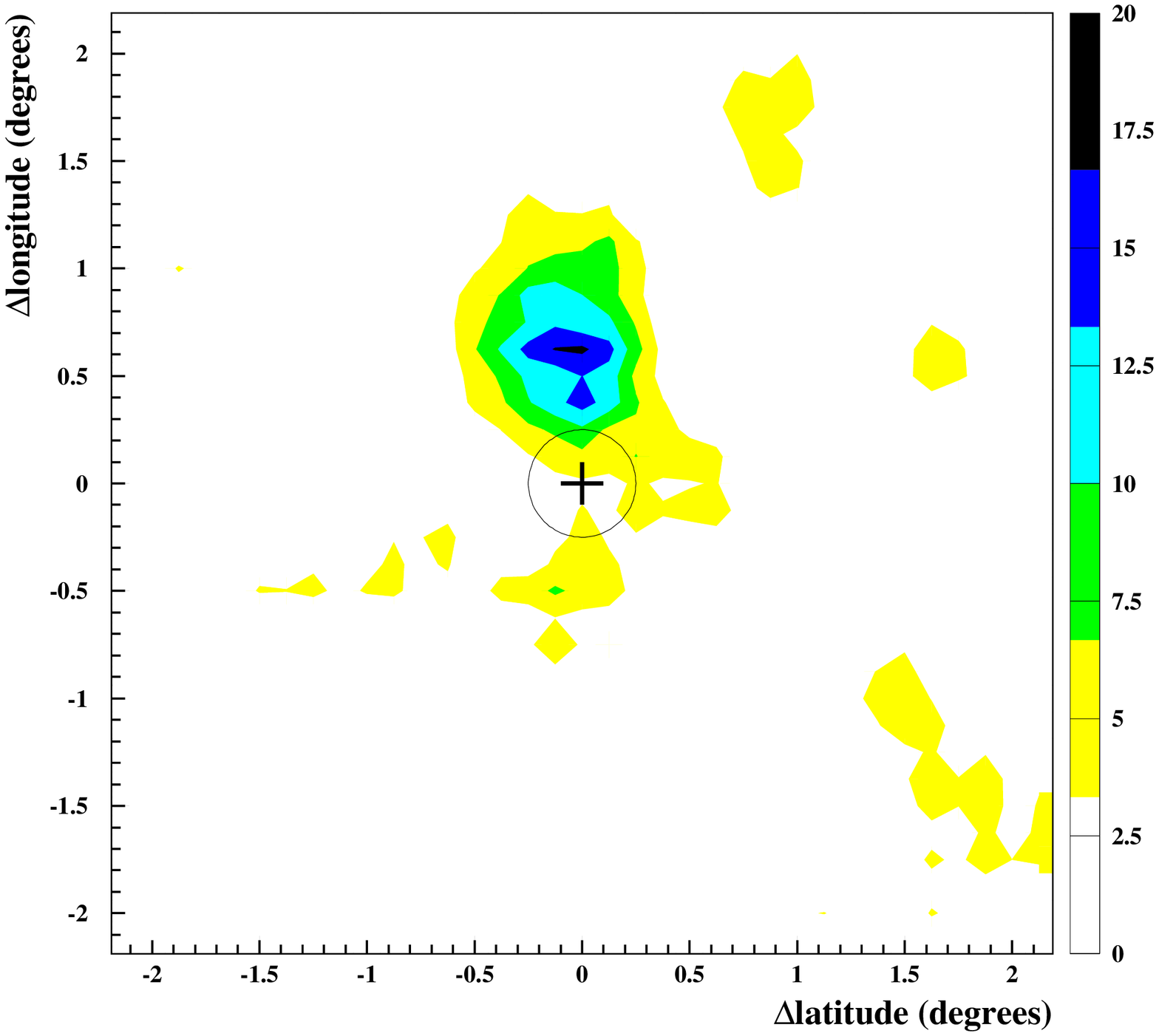}   }
 \end{center}
{\small \hskip 4.0 truecm (a) \hskip 7.5 truecm (b)}
\caption{\label{fig19}\footnotesize { 
Moon and sun shadows. (a) Two dimensional distributions of  muon event
density around the moon direction. The  regions of increasing gray
scale indicate increasing levels of deficit in percent.  
(b) Same analysis for the sun direction.}}
\end{figure}
\noindent \textit{Sidereal anisotropies}  are
due to the motion of the solar system in the  ``sea''  of relativistic cosmic rays
in our galaxy. They are expected to yield a small effect.  
After a correction due to the motion of the Earth around the Sun, we
observed variations with  an amplitude $8.6 \cdot 10^{-4}$ and a phase $\phi_{max} = 22.7^\circ$
with a statistical significance of 3 $\sigma$, Fig.  \ref{fig18}b.\\
\textbf{Moon and sun shadows of primary cosmic rays.} The pointing
capability of MACRO was demonstrated by the observed ``shadows''
of the Moon and of the Sun. A sample of \( 45\cdot 10^{6} \) muons was used to
look at the bidimensional density of the events around the directions
of the Moon and of the Sun,  Fig. \ref{fig19}. For the Moon: we looked for events in a window \( 4.375^{o}\times 4.375^{o} \) centered on the Moon; the window was divided into \( 35\times 35 \)
cells, each of dimensions  \( 0.125^{o}\times 0.125^{o} \)
(\( \Delta \Omega =1.6\cdot 10^{-2}deg^{2} \)); Fig. \ref{fig19}a shows a depletion of events with
a statistical significance of \( 5.5\, \, \sigma  \). The observed
slight displacement of the maximum deficit is consistent with the
displacement of the primary protons due to the geomagnetic field.
The same analysis was  repeated for muons in the sun window, Fig.
\ref{fig19}b. The  difference between the apparent sun position
and the observed muon depletion is due to the combined effect of the
magnetic field of the Sun and of the Earth. The observed
depletion has a statistical significance of \( 4.5\, \, \sigma  \)\cite{mac46}\cite{mac26}.\\
\textbf{Muon energy measurement with the TRD detector. }
The underground differential energy spectrum of muons was measured
with the  TRD  detector. Two types of
events were analyzed: ``single muons'', i.e. single muons in MACRO
crossing a TRD module, and ``double muons''
in MACRO with only one muon crossing the TRD detector. The
measurements refer to muons with energies 0.1$<$$E_{\mu }$$<$1$\TeV$ and for $ E_{\mu }$$>$1$\TeV$ \cite{mac27}\cite{mac44}. In order to evaluate
the local muon energy spectrum, we took  into account the TRD
response function.
The average single muon
energy in the  underground lab is \( 270\, \, \GeV  \);
for double muons  is  380 GeV. 
\section{EAS-TOP/MACRO Coincidences }
For coincident events, the Extensive Air Showers EASTOP  measured the e.m. size of the showers at Campo Imperatore, while MACRO measured  muons underground.
The purpose was the study of primary CR composition vs energy
reducing the dependence on the interaction and propagation models.
The two  detectors  operated in coincidence for a livetime of  960 days.
The number of coincident events was 28160, of which 3752
had shower cores inside the edges of  EASTOP  (``internal events'')
and shower sizes $N_e$$>$2$\cdot 10^5$;  409 events had $N_e$$>$$10^{5.92}$, i.e.
above the CR knee.
The data  were analyzed in terms of the QGSJET 
interaction model  implemented in CORSIKA \cite{mac14}\cite{mac49}\cite{mac50}\cite{kna5}. 

The e.m. detector of EASTOP  covered an effective area of  $10^5 \m^2$.
The array was fully efficient for $N_e$$>$$10^5$.
The  reconstruction capabilities of the  air shower  parameters 
for internal events were:
${{\Delta N_e} \over N_e} \simeq 10 \%$ for $N_e \aprge 10^5$,
and $\Delta \theta \sim 0.9^o$ for the  arrival direction. MACRO  considered  muons with $\geq$ 4 aligned hits in both views of the horizontal streamer tube planes.
Inside the EASTOP effective area, the muon energy threshold at the surface, for muons reaching  MACRO,
ranged from  1.3 to  1.8 TeV.   
Coincidences were made off-line, using the absolute time given by
a GPS system with an accuracy $\leq$1 $\mu s$.

The data considered were the muon multiplicity
distributions in  different intervals of shower sizes. For each 
bin the muon  distribution was fitted with a superposition of (i) pure $p$ and $Fe$
components, or (ii) light $(L)$ and
heavy $(H)$ admixtures containing equal fractions of $p$ and $He$ or $Mg$ and 
$Fe$, respectively.  
All spectra in the simulation have slope $\gamma = 2.62$. In each interval a $\chi^2$ expression was minimized.

\begin{figure}
  \begin{center}

  \mbox{ 
\epsfysize=8cm
         \epsffile{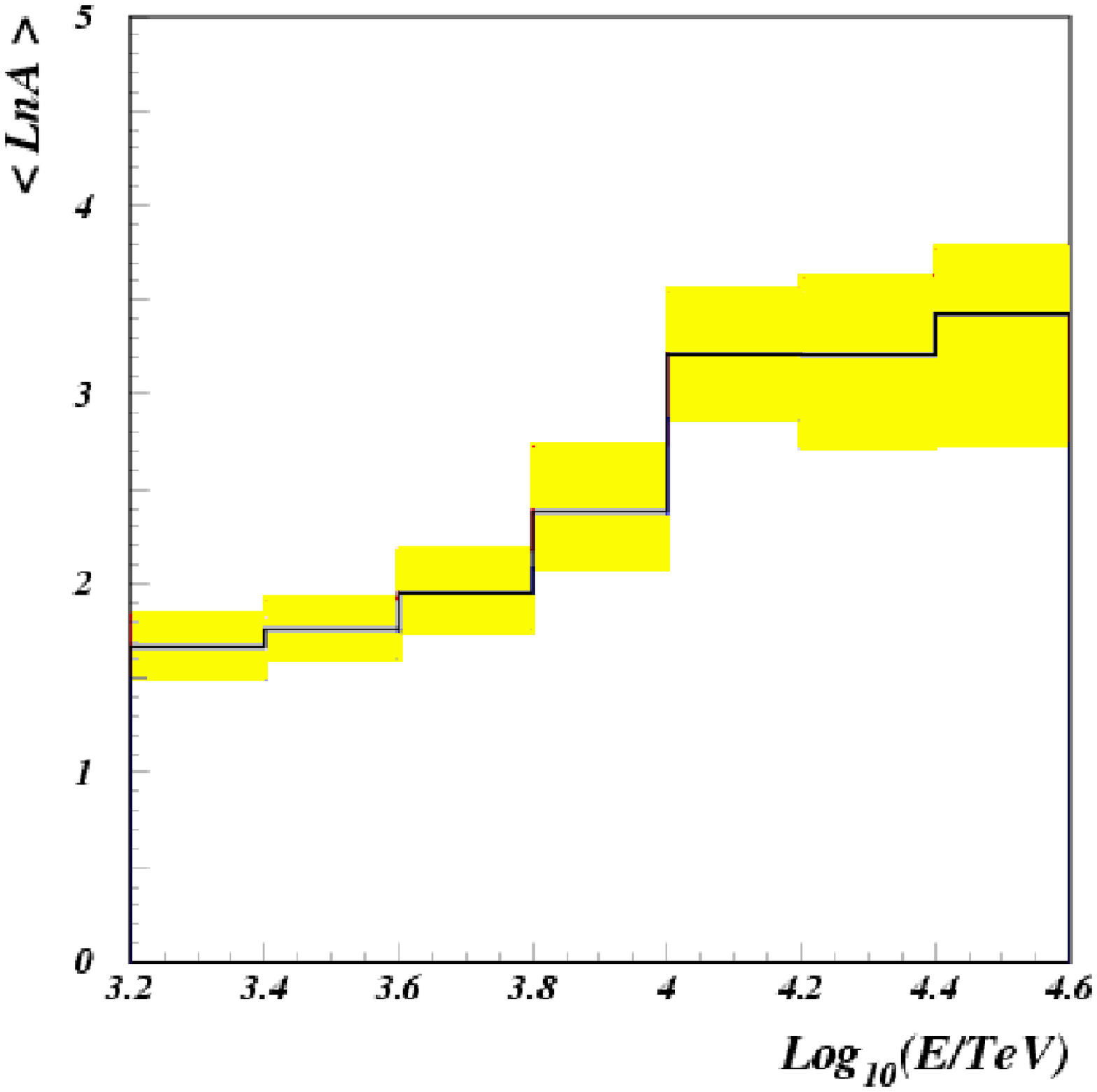} 
\epsfysize=8cm
         \epsffile{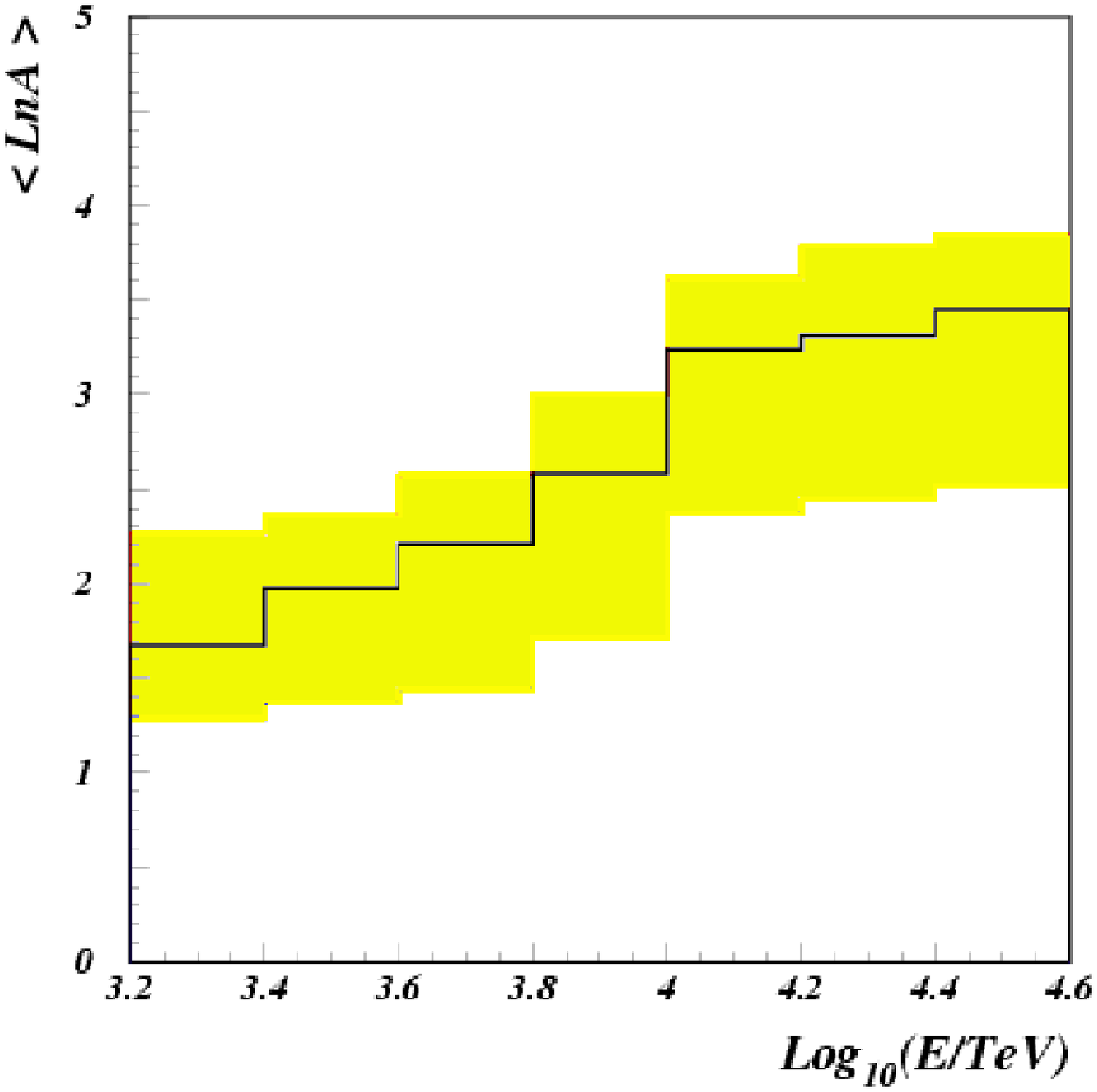} 
}   
{\small \hskip 5.0 truecm (a) \hskip 7.5 truecm (b)}
\caption{\label{fig21}\footnotesize {  EASTOP-MACRO coincidences. $\langle ln A
  \rangle$ vs primary energy for: (a) $p/Fe$  and (b) $Light/Heavy$ 
 compositions.
The (black lines) are the data, the shaded areas indicate the   uncertainties discussed in the text.
}}
\end{center}
\end{figure}

Fig. \ref{fig21} shows $\langle lnA \rangle$ 
vs  $log_{10}E$ (E in TeV); the shaded regions  include the uncertainties   
for (a) the p/Fe composition model and (b) for the light/heavy model. 
There is an increase of $\langle lnA \rangle$
with energy in the CR  knee region.

The coincidences allowed to 
measure the lateral distribution of Cherenkov light collected by  EASTOP in the 10$\div$100 TeV 
range  associating it with the TeV muon  in  MACRO. 
The test provided a validation of the CORSIKA-QGSJET code \cite{kna5}.

\section{Nuclear Track Detector Calibrations}

Many calibrations of the CR39 nuclear track detector (NTD) were performed
with both slow and fast ions. In all measurements we have seen no
deviation of its response from the Restricted Energy Loss (REL) model.
Stacks of
CR\( 39 \) and Lexan foils, placed before and after various targets,
were exposed to 158 A \GeV \( \, \, Pb^{82+} \) ions at the CERN-SPS
and to 1 A \GeV \( \, \, Fe^{26+} \) ions at the BNL-AGS. In traversing
a target, the beam ions produce nuclear fragments with $Z$$<$82 and $Z$$<$26 for the lead and iron beams, respectively; this allows
a measurement of the response of the detector in a \( Z \) region
relevant to   MM detection. Previous analyses had shown that the CR39 charge resolution for a single measurement is  $0.19 e$ in the range 
$72$$\le$Z$\le$ 83  (by measurements of the etch-cone
heights); at lower Z the measurements of the cone base diameters allow to separate the different charges. 
Tests were made looking for a possible aging effect, from the time elapsed from the production date
to the   exposure date \cite{miriam}. Two sets
of sheets, 0.8 y and 2.5 y old,  were exposed in 
1994 to 158 A \GeV \( \, \, Pb^{82+} \) ions. For each detected nuclear
fragment the reduced etch rate \( p=v_{T}/v_{B} \) (\( v_{T} \)
and \( v_{B} \) are the track and bulk etching rates)
was computed and plotted  vs REL. 
Within the experimental uncertainties, aging effects in the CR39 are negligible.

\section{Search for Lightly Ionizing Particles}

Free fractionally charged particles could be expected in Grand Unified Theories; the expected charges range 
from Q=\textit{e}/5  to Q=\textit{e} 2/3. They should release a fraction $(Q/e)^2$
of the energy deposited by a relativistic muon traversing a medium. Lightly
Ionizing Particles (LIPs) were searched for in MACRO using a four-fold
coincidence between three layers of scintillators and the streamer
tube system \cite{mac32}\cite{mac51}. The 90 \% c.l. flux upper limits for LIPs with
charges 2\textit{e}/3, 
\textit{e}/3  and \textit{e}/5 are  at the level of 
$1.5 \cdot 10^{-15} cm^{-2}
s^{-1}sr^{-1}$.
\section{Conclusions}
 MACRO obtained important results in all the items listed in the proposal :\\
- \textit{GUT Magnetic Monopoles.} We obtained the best flux upper limit over
the widest $\beta$ range, thanks to the large acceptance and the redundancy
of the different techniques employed. This limit  is a unique
result and it will stand for a long time.\\
- \textit{Atmospheric neutrino oscillations.} In this field MACRO  had its
major achievements. Analyses of different event topologies, different
energies and the exploitation of Coulomb multiple scattering in the detector
 strongly support  $\nu_\mu \rightarrow \nu_\tau$ oscillations.\\
- \textit{High energy $\nu_\mu$ astronomy.} Our detector was  competitive
with other underground/underwater experiments thanks to its good angular accuracy.
\\
- \textit{Search for bursts of $\bar{\nu}_e$ from stellar gravitational
collapses.} MACRO was sensitive to supernovae events in the Galaxy,
it started  the SN WATCH  system, and for a period it was the only detector in operation.\\
- \textit{Cosmic ray downgoing muons.} We  observed the shadows of primary CRs by the Moon and the
Sun, the seasonal variation ($\sim$2$\%$ amplitude) over many years, solar
and sidereal variations with reasonable statistical significances. No excesses of secondary muons attributable to astrophysical point sources (steady, modulated or bursting) were observed. \\
- Exploration of the \textit{CR composition} around
the {}``knee{}'' of the primary CR energy.\\
- \textit{Coincidences  between MACRO and the EASTOP array.} The data indicate an increase with increasing energy of the average Z of the primary CR nuclei.\\
- Sensitive searches for exotic particles were carried
out: (i) \textit{WIMPs}, looking
for upgoing muons  from the center of the Earth and of the Sun;
(ii) \textit{Nuclearites} and \textit{Q-balls} (as byproducts of MM searches).
(iii) Other limits concern possible \textit{Lightly Ionizing Particles}.\\

The dismantling of MACRO went regularly and essentially on schedule.
We recuperated part of the electronics (modules, circuits, cables,
etc) to be used in our Institutions, and donated the photomultipliers
and part of the streamer tubes to other experiments.

The MACRO scientific and technical results were  published in 50 papers in refereed journals, in  242 contributions to  conferences and in invited papers, discussed in  543 Internal Memos, used for 83 italian Laurea theses, 22 italian Dottorato theses, 23 US PhD theses and 5 moroccan theses de Doctorat Nationale. 
\subsection*{Acknowledgments}
{\small{ We thank all our colleagues for their cooperation, and wish all the best to Charles Peck.\\
MACRO acknowledged the support of the Directors and of the staff of the
Gran Sasso Lab. and of the Institutions partecipating in the
experiment, the Istituto Nazionale di Fisica Nucleare (INFN),
the US Department of Energy and the US National Science Foundation. INFN-FAI, ICTP (Trieste), NATO and WorldLab  provided
fellowships and grants for non-Italian citizens.}}

\end{document}